%% file: main_v2.tex
\documentclass{SciPost}

\hypersetup{
    colorlinks,
    linkcolor={red!50!black},
    citecolor={blue!50!black},
    urlcolor={blue!80!black}
}

\usepackage[bitstream-charter]{mathdesign}
\DeclareSymbolFont{usualmathcal}{OMS}{cmsy}{m}{n}
\DeclareSymbolFontAlphabet{\mathcal}{usualmathcal}

\fancypagestyle{SPstyle}{
\fancyhf{}
\lhead{\colorbox{scipostblue}{\bf \color{white} ~SciPost Physics Codebases }}
\rhead{{\bf \color{scipostdeepblue} ~Submission }}

\fancyfoot[C]{\textbf{\thepage}}
}

\usepackage{graphicx} 
\usepackage{physics}
\usepackage{algorithm}
\usepackage{algpseudocode}
\usepackage{mathtools}
\usepackage{amsmath}
\usepackage{dsfont}

\usepackage{amssymb}
\usepackage{booktabs}

\usepackage{hyperref}
\usepackage{wrapfig}
\usepackage{tikz}
\usetikzlibrary{shapes.geometric, arrows}
\usetikzlibrary{positioning}
\usepackage[utf8]{inputenc}

\usepackage{orcidlink}

\usepackage{xcolor}
\usepackage[normalem]{ulem}
\definecolor{darkgreen}{rgb}{0,0.60,.2}
\definecolor{darkblue}{rgb}{0.1,0.3,1}

\newcommand{\tl}{\tilde{\lambda}}
\newcommand{\tmap}{t_{\mathrm{mapping}}}
\newcommand{\tdiag}{t_{\mathrm{diag}}}
\newcommand{\tzm}{t_0^{\mathrm{mapp}}}
\newcommand{\tzd}{t_0^{\mathrm{diag}}}
\newcommand{\ddd}{\mathrm{d}}

\newcommand{\inh}{\inlinenohighlight}

\tikzstyle{arrow} = [thick,->,>=stealth, very thick]
\tikzstyle{arrowphases} = [thick,->,>=stealth,  solid]
\tikzstyle{inputoutput} = [rectangle, rounded corners, minimum width=3cm, minimum height=1cm,text centered, draw=black, outer color=blue!30, inner color = white]
\tikzstyle{unit} = [rectangle, rounded corners, minimum width=3cm, minimum height=1cm,text centered, draw=black, outer color=green!30, inner color = white]
\tikzstyle{decision} = [diamond, rounded corners, minimum width=3cm, minimum height=1cm,text centered, draw=black, outer color=orange!30, inner color = white]
\tikzstyle{phase} = [rectangle, rounded corners, minimum width=3cm, minimum height=1cm,text centered, draw=black, dashed, fill=yellow!30]

\newcommand{\polfedjl}{\href{https://github.com/RockClimbingRocks/Polfed.jl}{\textbf{Polfed.jl}}}

\input{jlcodeblock_setup.tex}

\begin{document}

\pagestyle{SPstyle}

\begin{center}{\Large \textbf{\color{scipostdeepblue}{
Computing eigenpairs of quantum many-body systems with Polfed.jl
}}}\end{center}

\begin{center}\textbf{
Rok Pintar\textsuperscript{1,2}\,\orcidlink{0009-0000-8491-3592}\,,
Konrad Pawlik\textsuperscript{3,4}\,\orcidlink{0009-0009-6304-2578}\,,
Rafał Świętek\textsuperscript{1,2,6},\,\orcidlink{0009-0004-5353-9998}\,,
Miroslav Hopjan\textsuperscript{1,7},\,\orcidlink{0000-0002-0905-1571}\,,
Jan \v{S}untajs\textsuperscript{2, 9}\,\orcidlink{0000-0003-1290-4923}\,,
Jakub Zakrzewski\textsuperscript{4,5}\,\orcidlink{0000-0003-0998-9460}\,,
Piotr Sierant\textsuperscript{8}\,\orcidlink{0000-0001-9219-7274}, and
Lev Vidmar\textsuperscript{1,2}\,\orcidlink{0000-0002-6641-6653} 
}\end{center}

\begin{center}
{\bf 1} Department of Physics, Faculty of Mathematics and Physics, University of Ljubljana, SI-1000 Ljubljana, Slovenia
\\
{\bf 2} Department of Theoretical Physics, Jo\v{z}ef Stefan Institute, SI-1000 Ljubljana, Slovenia
\\
{\bf 3} Szkoła Doktorska Nauk Ścisłych i Przyrodniczych, Uniwersytet Jagielloński, ulica Stanisława \L{}ojasiewicza 11, PL-30-348 Krak\'ow, Poland

{\bf 4} Instytut Fizyki Teoretycznej, Wydzia{\l} Fizyki, Astronomii i Informatyki Stosowanej, Uniwersytet Jagiello\'{n}ski, {\L}ojasiewicza 11, PL-30-348 Krak\'{o}w, Poland
\\
{\bf 5} Mark Kac Complex Systems Research Center, Jagiellonian University in Krak\'{o}w, PL-30-348 Krak\'{o}w, Poland
\\
{\bf 6} Institut f\"ur Theoretische Physik, Georg-August-Universität G\"ottingen, D-37077 G\"ottingen, Germany
\\
{\bf 7} Institute of Theoretical Physics, Faculty of Fundamental Problems of Technology, Wrocław University of Science and Technology, 50-370 Wrocław
\\
{\bf 8} Barcelona Supercomputing Center, Pla\c{c}a Eusebi G\"{u}ell 1-3, 08034 Barcelona, Spain 
\\
{\bf 9}  Faculty of Mechanical Engineering, University of Ljubljana, SI-1000 Ljubljana, Slovenia
\\
\end{center}

\section*{\color{scipostdeepblue}{Abstract}}
\textbf{\boldmath{%
We present \polfedjl, an open-source Julia package implementing the Polynomially Filtered Exact Diagonalization (POLFED) algorithm for computing mid-spectrum eigenvalues and eigenvectors (shortly, eigenpairs) of quantum many-body Hamiltonians. Access to such eigenpairs is essential for studying non-equilibrium many-body physics, but is hindered by the exponential growth of Hilbert-space dimension. POLFED addresses this challenge through a polynomial spectral transformation evaluated on the fly within a Lanczos iteration, preserving Hamiltonian sparsity and substantially reducing memory costs compared to other diagonalization methods. The package supports flexible energy targeting, automatic optimization of the spectral mapping for structured Hamiltonians, and GPU acceleration, which is particularly effective since the dominant computational cost reduces to repeated sparse matrix-vector multiplications. Benchmarks on disordered spin chains and fermionic models demonstrate access to larger system sizes than alternative approaches, and CPU--GPU comparisons confirm significant speedups. In particular, we also provide code for constructing the quantum sun model Hamiltonian, a toy model of a many-body ergodicity-breaking transition. While our focus is on many-body Hamiltonians, \inh{Polfed.jl} may be applied to any large sparse matrix.
}}

\vspace{\baselineskip}

\noindent\textcolor{white!90!black}{%
\fbox{\parbox{0.975\linewidth}{%
\textcolor{white!40!black}{\begin{tabular}{lr}%
  \begin{minipage}{0.6\textwidth}%
    {\small Copyright attribution to authors. \newline
    This work is a submission to SciPost Physics. \newline
    License information to appear upon publication. \newline
    Publication information to appear upon publication.}
  \end{minipage} & \begin{minipage}{0.4\textwidth}
    {\small Received Date \newline Accepted Date \newline Published Date}%
  \end{minipage}
\end{tabular}}
}}
}


\vspace{10pt}
\noindent\rule{\textwidth}{1pt}
\tableofcontents
\noindent\rule{\textwidth}{1pt}
\vspace{10pt}


\section{Introduction}
\label{sec:intro}
Over the past hundred years of quantum mechanics development and application, the exact diagonalization has remained a prime {\it tour de force} numerical approach in the treatment of various problems from simple, single-particle models, to many-body problems~\cite{Lin1990, Laflorencie2004, Sandvik2010, QuSpin2017, Lami2025, XDiag2026}. A successful diagonalization of the Hamiltonian, yielding eigenvalues and eigenvectors, is a prerequisite for an analysis of both static and dynamic properties. While in this work we shall mainly consider the interacting low-dimensional systems, the methods discussed may easily find application in other areas where a partial diagonalization of large, sparse matrices forms an essential step in the treatment of a given problem.

The study of quantum phenomena within lattice geometries commenced in the early 1920s. Despite the existence of the Bethe Ansatz \cite{Bethe1931} for solving the one-dimensional antiferromagnetic isotropic (XXX) Heisenberg model \cite{Lieb63a, Lieb63b, Yang66a, Yang66b, Lieb68, Gaudin_2014}, there are very few analytical solutions. The lack of comprehensive analytical results prompted a shift towards numerical simulations, especially with the rapid advancement of computational technology. One of the earliest numerical methods employed was exact diagonalization (ED), i.e., the computation of all eigenvalues and eigenvectors of the Hamiltonian matrix using standard dense linear algebra routines~\cite{LAPACK1999}, often regarded as a brute-force approach. Although ED provides access to eigenvalues and eigenfunctions, allowing for the calculation of various quantities, its applicability is limited by the exponential growth of the Hilbert-space dimension with the system size and the cubic scaling of the diagonalization cost with the matrix dimension. Consequently, it becomes impractical for studying large quantum systems.

Over time, numerous methods have evolved to address larger systems, albeit often with specific constraints. For instance, some methods are suitable only for low or high-temperature properties, or they can only provide thermodynamic properties without yielding eigenvalues. Frequently, researchers are primarily interested in the low-lying eigenstates or even just the ground state. The development of the Lanczos algorithm \cite{Lanczos1950} marked a significant breakthrough in this regard. This method gained prominence with the increase in computational power and saw further advancements \cite{PhysRevB.Continued-fraction, Benoit_1992, Haydock_1972, PhysRevB.Low-temperature-Lanczos-method, PhysRevB.finite-temperature-Lanczos}, and a variety of applications \cite{José-M-Soler_2002, RevModPhys-electronic-structure–of-impurities, jaklic-prelovsek-2000, RevModPhys.Correlated-electrons-in-high-temperature-superconductors}. The Lanczos method is now widely employed to calculate dynamical and thermodynamic properties of strongly correlated systems at both low and high temperatures. Similar quantities can be obtained using the Kernel Polynomial Method (KPM) \cite{kpm}, which offers certain advantages over the Lanczos algorithm but does not yield explicit eigenvalues and eigenvectors.

In the study of thermalization \cite{Deutsch1991, Srednicki1994, Polkovnikov11review, DAlessio16review} and ergodicity-breaking phenomena \cite{Nandkishore15mbl, Abanin19colloquium, Laflorencie16entanglement, Alet18, Sierant25mblreview, Serbyn21scars, Chandran23scars, Moudgalya22fragmentation} in quantum many-body systems, access to individual eigenvalues and eigenstates is essential. To mitigate finite-size effects when analyzing the behavior of ergodicity indicators, only the central part of the spectrum, i.e., high-temperature eigenvalues and eigenstates, is considered. Due to the nature of the Lanczos algorithm, obtaining the eigenpairs in the bulk of the spectrum necessitates transforming the Hamiltonian to expose the desired spectral region. 
A major step forward was the shift-and-invert method~\cite{Pietracaprina2018}, which makes mid-spectrum eigenpairs accessible to Krylov solvers by spectral inversion. However, it requires a factorization of the shifted Hamiltonian, leading to memory requirements that prohibit reaching larger system sizes.
This motivated the development of the Polynomially Filtered Exact Diagonalization (POLFED) method~\cite{Sierant20polfed}. While building on the general idea of polynomial filtering eigensolvers~\cite{Saad11eigenvalue, Fang12filteredLanczos, Pieper16chebfd}, POLFED introduces key adaptations tailored to quantum many-body Hamiltonians, in particular an on-the-fly spectral transformation that preserves sparsity throughout the calculation.

In this paper we present \inh{Polfed.jl}, an open-source Julia implementation of the POLFED algorithm capable of handling both real symmetric and general complex hermitian matrices, and provide a pedagogical introduction to the underlying methods.
The article is structured as follows. We first review Krylov space methods essential for the Lanczos algorithm and its block generalization in Sec.~\ref{sec:krylov:space:methods}. We then introduce spectral transformations that allow one to address a desired part of the spectrum in Sec.~\ref{Methods:for:targeting:central:part:of:the:spectrum}. The details of the POLFED algorithm are presented in Sec.~\ref{sec:POLFED}, followed by the description of the \inh{Polfed.jl} interface in Sec.~\ref{sec:interface}. Benchmarks are reported in Sec.~\ref{sec:benchmarks}. We conclude in Sec.~\ref{sec:conclusion}. Appendices contain filtered-subspace diagonalization algorithms, numerical details about convergence of eigenpairs, scaling of the polynomial order, a guide to choosing the number of requested eigenpairs for optimal performance, details of the models studied---including the quantum sun model, for which we provide the code for its implementation---and describe the Kernel Polynomial Method.

\section{Krylov space methods}
\label{sec:krylov:space:methods}

In the absence of analytical results, researchers have resorted to numerical simulations to study physical systems. Perhaps one of the simplest methods is exact diagonalization (ED), which is the only method granting access to the full energy spectrum and the corresponding eigenstates. As such, ED is an invaluable tool in different fields of study, such as the eigenstate thermalization hypothesis~\cite{Deutsch1991, Srednicki1994, DAlessio16review} and ergodicity breaking transitions~\cite{Nandkishore15mbl, Alet18, Abanin19colloquium, Sierant25mblreview}. The exponential growth of the many-body Hilbert-space dimension $\mathcal{D}$ with the system size $L$, however, limits its applicability to relatively small systems. 
    
In typical systems of interest, such as spin-1/2 chain or spinless fermionic systems on $L$ lattice sites, one has full Hilbert space $\mathcal{D}=2^L$ with no symmetries. Since, to perform ED, one needs to diagonalize a matrix of dimensions $\mathcal{D} \times \mathcal{D},$ the calculation quickly becomes challenging upon increasing $L$, both in terms of memory consumption and computational complexity, which scales as $\mathcal{O}( \mathcal{D}^3)$~\cite{GolubVanLoan2013}. Additionally, ED does not exploit the sparsity of typical quantum many-body Hamiltonians. For systems with local or short-range interactions, each basis state is coupled only to few others~\cite{Sandvik2010, avella2013}, so that the number of nonzero matrix elements per row typically scales $\propto L,$ while the number of all possible states scales $\propto 2^L,$ resulting in a very sparse matrix. While this sparsity could be efficiently exploited by storing the matrix in one of the many available sparse formats and using it for sparse matrix-vector multiplication, ED ultimately requires a fully dense matrix. To exploit sparsity efficiently, one typically resorts to iterative methods, such as the Lanczos algorithm, which we explain in more detail below.

\subsection{Introduction to Lanczos algorithm}
\label{sec:lanczos}
The Lanczos algorithm~\cite{Lanczos1950} is an iterative method for computing extremal eigenvalues and eigenvectors of large sparse symmetric (hermitian) matrices~\cite{GolubVanLoan2013, Saad11eigenvalue}. We denote  the number of Lanczos iterations as $m$, which corresponds to a Krylov basis of dimension $m$. For the conventional Lanczos algorithm, we simply use $m$ to refer to the number of iterations. However, the iteration count and the total basis dimension scale differently in the block Lanczos algorithm introduced in Sec.~\ref{sec:block:lanczos}, where we will use $M$ to strictly denote the total Krylov space dimension. Both Lanczos and its block variant belong to the family of Krylov-subspace methods, which approximate the solution within the subspace:
\begin{equation} 
      \mathcal{K}_m(H,\ket{\phi_1}) = \mathrm{span}\{\ket{\phi_1}, H\ket{\phi_1}, H^2\ket{\phi_1}, \dots, H^{m-1}\ket{\phi_1}\}\,,
      \label{eq:KrylovSubspace}
 \end{equation}
generated by repeated application of the matrix $H$ to an initial vector $\ket{\phi_1}$. Since its introduction, many variants of the Lanczos algorithm have been developed, including the implicitly restarted Lanczos method~\cite{Sorensen92implicitRestart, Lehoucq98arpack}, which controls memory usage and loss of orthogonality by periodically restarting the iteration with an improved starting vector, the finite-temperature Lanczos method~\cite{jaklic-prelovsek-2000}, and the block Lanczos method~\cite{Golub77blockLanczos}, discussed in the following sections.

The idea of the Lanczos algorithm is as follows. Starting from a random normalized vector $\ket{\phi_1}$\footnote{We assume $\ket{\phi_1}$ has nonzero overlap with each target eigenstate of $H$.},  we iteratively construct a tridiagonal representation of $H$ by building an orthonormal basis of the Krylov subspace~\eqref{eq:KrylovSubspace}. In the first step, we apply $H$ to the initial vector and decompose the result into components parallel and orthogonal to $\ket{\phi_1}$:
\begin{align}
    \label{eq:lanczos_firststep}
    H\ket{\phi_1} = a_1 \ket{\phi_1} + b_2 \ket{\phi_2}.  
\end{align}
The coefficients are
$a_1=\bra{\phi_1}H\ket{\phi_1}$ and $b_2=\bra{\phi_2}H\ket{\phi_1}$. Hermiticity of $H$ ensures that $a_1$ is real, while the phase of $\ket{\phi_2}$ can always be chosen so that $b_2$ is real as well. In the next step, we apply $H$ to $\ket{\phi_2}$:
        \begin{align}
            H\ket{\phi_2} = b_2^\prime \ket{\phi_1} + a_2 \ket{\phi_2} + b_{3} \ket{\phi_3}.     
        \end{align}
Here, $b_2^\prime = \bra{\phi_1}H\ket{\phi_2} = b_2$ and $\ket{\phi_3}$ is orthogonal to both $\ket{\phi_1}$ and $\ket{\phi_2}$. Continuing this procedure, at the $i$-th step we obtain:
\begin{align}
    \label{eq:lanczos_recurrence}
    H\ket{\phi_i} = b_{i} \ket{\phi_{i-1}} + a_{i} \ket{\phi_i} + b_{i+1} \ket{\phi_{i+1}}, \hspace{5mm} 1 < i \leq m.
\end{align}
We truncate the iteration by setting $b_{m+1}=0$. The above procedure generates an orthonormal basis of Krylov vectors, $\kappa = \{\ket{\phi_1}, \ket{\phi_2}, \dots, \ket{\phi_m}\},$ which defines a projection of the full Hamiltonian onto the $m$-dimensional Krylov subspace. This projection yields the \emph{Lanczos matrix}, $H_L \in \mathbb{R}^{m \times m}.$ By construction, $H_L$ is a tridiagonal matrix with diagonal elements $a_i$ and off-diagonal elements $b_i$ on the leading sub- and superdiagonals, which in the case of the standard Lanczos algorithm can be made real, by absorbing complex phases into $\ket{\phi_i}$.
Compared to ED, diagonalization of tridiagonal matrices is much faster, since the cost scales as $\mathcal{O}(m^2)$~\cite{GolubVanLoan2013}, and several specialized routines are readily available for the task. Let $t_j = (t_{1j}, t_{2j}, \dots, t_{mj})^T$ be the exact orthonormal eigenvectors of $H_L$ with the corresponding eigenvalues $\varepsilon_j$. Because $H_L$ is the representation of $H$ restricted to the Krylov subspace, we construct the approximate eigenvectors of the full Hamiltonian $H$, known as Ritz vectors $\ket{\psi_j}$, by taking a linear combination of the Krylov basis vectors weighted by the scalar components of $t_j$,
\begin{equation}\label{eq:lan_eigvecs}
\ket{\psi_j} = \sum_{i=1}^m t_{ij}\ket{\phi_i}.
\end{equation}
The corresponding Ritz values, which approximate the exact eigenvalues of $H$, are the eigenvalues of $H_L$, and can be equivalently written as the expectation value of the full Hamiltonian: $\varepsilon_j=\mel{\psi_j}{H}{\psi_j}$. In general, $\ket{\psi_j}$ do not coincide with the exact eigenvectors $\ket{u_j}$ of $H,$ since $H_L$ is typically only an approximation of the Hamiltonian. The exception occurs when $b_{m+1}=0$ in Eq.~\eqref{eq:lanczos_recurrence} is satisfied exactly, indicating that $\kappa$ is an $m$-dimensional invariant subspace, in which case $H_L$ exactly represents $H$. This situation arises trivially when $m = \mathcal{D}$, but it can also occur for $m<\mathcal{D}$ if the starting vector $\ket{\phi_1}$ is orthogonal to some invariant subspace of $H$\footnote{In practice, this is avoided by choosing $\ket{\phi_1}$ as a random vector, since the probability of a randomly sampled vector lying exactly within a proper invariant subspace of $H$ is vanishingly small.}. In all other cases, setting $b_{m+1}=0$ introduces a truncation error. In practice, sufficient convergence is typically reached well before $m$ reaches $\mathcal{D}$ and the corresponding truncation error can be estimated \cite{Parlett1998SymmetricEigenvalueProblem} as
\begin{align}\label{eq:lan_criterion}
    H\ket{\psi_j} - \varepsilon_j\ket{\psi_j} = b_{m+1} t_{mj} \ket{\phi_{m+1}}.
\end{align}
A natural convergence criterion is $|b_{m+1} t_{mj}| < \epsilon$, for some threshold value $\epsilon > 0$.

The Lanczos method is particularly suitable for finding extremal eigenvalues. This selective convergence occurs because the projection onto the Krylov subspace naturally isolates extreme energy states while suppressing the densely packed states in the middle of the spectrum. We rigorously detail the bounds of this convergence mechanism in Sec.~\ref{sec:convergence:krylov:methods}.
In practical applications, the typical number of iterations required for convergence satisfies $50 \lesssim m \lesssim 100$, with the total cost (without reorthogonalization, see below) scaling as $\mathcal{O}(m\, n_{\rm{nz}} \mathcal{D})$ where $n_{\rm{nz}}$ represents the number of nonzero off-diagonal matrix elements per row.

Another important practical issue is the loss of orthogonality among the Lanczos vectors $\ket{\phi_i}$. After sufficiently many iterations, the accumulation of errors due to finite precision leads to a noticeable degradation of orthogonality. This is particularly pronounced when targeting eigenvalues with a small spectral gap. To restore orthogonality, one must carry out reorthogonalization at each Lanczos step, which requires storage of all previously computed Krylov vectors. This increases the cost to $\mathcal{O}( m^2 \mathcal{D})$. In practice, the cost can be reduced by selective~\cite{Parlett79selective} or partial~\cite{Simon84partial} reorthogonalization, which carries out reorthogonalization only when the overlap among Lanczos vectors exceeds a prescribed threshold. In the block Lanczos variant used by POLFED, however, full reorthogonalization is employed to ensure robust convergence for mid-spectrum eigenvalues.

\subsection{Generalization to block Lanczos method}
\label{sec:block:lanczos}
The standard Lanczos method, as described above, is known to be poorly parallelizable: each iteration depends sequentially on the previous one, and the single matrix-vector multiplication in Eq.~\eqref{eq:lanczos_recurrence} is the main step that can benefit from parallelism. A natural way to increase the arithmetic intensity and improve parallel efficiency is to operate on multiple vectors simultaneously, which is the idea behind the \emph{block Lanczos method}~\cite{Golub77blockLanczos}.
Instead of a single matrix-vector product per iteration, the block Lanczos method performs matrix-matrix multiplications between the Hamiltonian and blocks of orthonormal vectors. This replaces BLAS-2 operations with BLAS-3 operations\footnote{BLAS (Basic Linear Algebra Subprograms) are standard low-level routines for linear algebra operations. BLAS-2 refers to matrix-vector operations, such as multiplying a matrix by a single vector, whereas BLAS-3 refers to matrix-matrix operations, such as multiplying a matrix by a block of vectors. BLAS-3 operations typically achieve better performance on modern hardware because they reuse data more efficiently and better exploit cache hierarchy and parallelism.}, which offer significantly better cache utilization\footnote{Cache utilization refers to how effectively a program reuses data stored in the processor's fast cache memory, instead of repeatedly loading it from slower main memory. Good cache utilization improves performance by organizing computations so that recently accessed data are reused before being evicted from cache.} and parallel scaling. In addition, the block approach improves convergence for clustered or degenerate eigenvalues~\cite{Golub77blockLanczos}.

We generalize a single starting vector to a block of $s$ orthonormal vectors:
    \begin{align}
        \ket{\phi_1} \in \mathbb{C}^{\mathcal{D}} \to V_1 \in \mathbb{C}^{\mathcal{D}\times s}.
    \end{align}
A simple and efficient way to ensure orthonormality of the initial vectors is via the QR decomposition of a $\mathcal{D}\times s$ random matrix. In analogy with the standard Lanczos method, the goal is to iteratively construct a \emph{block tridiagonal} Lanczos matrix $H_L \in \mathbb{C}^{M \times M}$ (here $M=ms$). At the $k$-th step, where $1 \leq k \leq m,$ we perform a series of operations. First, we multiply the current block by the Hamiltonian:
    \begin{equation}
        W_k = H \, V_k.
    \end{equation}
Next, we project $W_k$ onto $V_k$, which defines the $k$-th diagonal block of $H_L$. Adopting a programmatic slicing notation where $a:b$ denotes the range of indices from a to b inclusive, we write:
    \begin{align}
        & A_k= V_k^\dagger W_k = V_k^\dagger H V_k, \\
        & H_L[(k-1)s+1 : ks, \, (k-1)s+1 : ks] = A_k.
    \end{align}
To prepare the next Krylov block, $W_k$ must be orthogonalized against all previously computed Krylov blocks $V_1, \dots, V_k$ (in exact arithmetic only $V_{k-1}$ and $V_k$ would be needed, but full reorthogonalization is necessary to maintain numerical orthogonality):
    \begin{equation}
        W_k \coloneqq W_k - \sum\limits_{i=1}^{k} V_i \left(V_i^\dagger W_k\right).
    \end{equation}
We then perform a QR decomposition of the orthogonalized block, which defines the next Krylov block and the next off-diagonal block:
    \begin{equation}
        W_k = V_{k+1}B_{k+1}, \hspace{5mm} V_{k+1} \in \mathbb{C}^{\mathcal{D} \times s}, \hspace{5mm} B_{k+1} \in \mathbb{C}^{s \times s}.
    \end{equation}
By definition of the QR decomposition, the columns of $V_{k+1}$ are orthonormal while $B_{k+1}$ is upper triangular.
For $k<m$, the off-diagonal blocks of $H_L$ are assigned as:
    \begin{align}
        & H_L[(k-1)s+1 : ks, ks+1 : (k+1) s]=B_{k+1}^\dagger,\\
        & H_L[ks+1 : (k+1) s, (k-1)s+1 : ks]=B_{k+1}.
    \end{align}
With this convention, and defining $V_0 = 0$ and $B_1 = 0$, the block Lanczos recurrence reads
    \begin{equation}
        \label{eq:block_lanczos_reocurrence}
        HV_k = V_{k-1} B_k^\dagger + V_k A_k + V_{k+1} B_{k+1}, \quad 1 \leq k \leq m.
    \end{equation}

The convergence criterion after $m$ block steps follows by strict analogy with the standard Lanczos method. Let $t_j$ be an orthonormal eigenvector of $H_L$ (the Lanczos matrix after $m$ block steps of dimension $M$). The Ritz vectors $\ket{\psi_j}$ in the full space are constructed as a linear combination of the block basis vectors: $\ket{\psi_j} = \sum_{i=1}^m V_i \tau_{ij}$, where $\tau_{ij}$ is the $i$-th block of size $s$ of the eigenvector $t_j$. By applying the block recurrence, the residual of the $j$-th Ritz pair evaluates to
\begin{align}
    H\ket{\psi_j} - \varepsilon_j\ket{\psi_j} = V_{m+1} B_{m+1} \tau_{mj}.
\end{align}
Since the columns of $V_{m+1}$ are orthonormal, the residual norm is simply $\|B_{m+1} \tau_{mj}\|$, with  $\|\bullet\|$ denoting the standard $\mathrm{L}^2$ norm. Thus, a natural block-Lanczos convergence criterion is
\begin{equation}\label{eq:blocklan_criterion}
    \max_j \norm{B_{m+1} \tau_{mj}} < \epsilon,
\end{equation}
where $\epsilon$ is the desired accuracy threshold. This is the direct block analog of Eq.~\eqref{eq:lan_criterion}.

\subsection{Convergence of Krylov-based methods}
\label{sec:convergence:krylov:methods}

As outlined in the introduction, the Lanczos algorithm selectively converges extremal eigenvalues within the Krylov subspace $\mathcal{K}_m(H,\ket{\phi_1})$ defined in Eq.~\eqref{eq:KrylovSubspace}. Specifically, because any vector in this subspace takes the form $P^{m-1}(H)\ket{\phi_1}$ for a polynomial $P^{m-1}$ of degree $m-1$, the Lanczos algorithm may be alternatively formulated as a search for an optimal $P^{m-1}$ that minimizes the approximation error for a given target state. As we demonstrate in this section, this optimization heavily accelerates convergence for the isolated extremal eigenvalues while suppressing it in the dense interior spectrum of a typical physical Hamiltonian. The rigorous mathematical foundation for this behavior is provided by the Kaniel-Paige convergence theory~\cite{Saad1980block, Parlett1998SymmetricEigenvalueProblem, Saad11eigenvalue, GolubVanLoan2013}, which bounds this approximation error using Chebyshev polynomials.

Let the exact eigenvalues of the hermitian matrix $H$ be ordered as $E_1 < E_2 < \dots < E_\mathcal{D}$. For now, we assume no degeneracies, as the standard Lanczos algorithm is not well suited for degenerate states~\cite{Golub77blockLanczos}; this limitation is removed in the block Lanczos algorithm, assuming exact arithmetic. Let $\varepsilon_1$ be the smallest eigenvalue of the Lanczos matrix $H_L$ at iteration $m$. The Kaniel-Paige theorem bounds the error of this approximation as~\cite{Saad1980block, Parlett1998SymmetricEigenvalueProblem, Saad11eigenvalue}
\begin{equation}
    0 \leq \varepsilon_1 - E_1 \leq  \frac{(E_\mathcal{D} - E_1)\tan^2 \theta_1}{[T_{m-1}(1 + 2\gamma_1)]^2}\,,
    \label{eq:kaniel_paige}
\end{equation}
where $\theta_1$ is the principal angle between the initial starting vector $\ket{\phi_1}$ and the exact eigenvector $\ket{u_1}$ corresponding to $E_1$, given by $\theta_1=\arccos |\braket{\phi_1}{u_1}|$. We notice that the numerator does not change with the number of Lanczos iterations $m$; thus, it serves as a fixed limitation on the convergence speed. The denominator is driven by $T_{m-1}$, the Chebyshev polynomial of the first kind of degree $m-1$, evaluated at $1 + 2\gamma_1$. The critical parameter governing convergence is the relative spectral gap, defined for the ground state as
\begin{equation}
    \gamma_1 = \frac{E_2 - E_1}{E_\mathcal{D} - E_2}\,.
\end{equation}
This gap measures the isolation of the target state ($E_1$) relative to the entire spread of the remaining spectrum ($E_2$ to $E_\mathcal{D}$).

The mechanism of convergence relies entirely on the properties of Chebyshev polynomials. For arguments $x \in [-1, 1]$, the polynomial is bounded such that $|T_{m-1}(x)| \leq 1$. However, for $x > 1$, the polynomial $T_{m-1}(x)$ grows exponentially with $m$, as will be demonstrated shortly. When computing the ground state, the Lanczos algorithm effectively maps the remaining spectrum into the bounded interval $[-1, 1]$, while the target extremal eigenvalue is mapped to $1 + 2\gamma_1$, where $\gamma_1>0$. To demonstrate this exponential convergence, we can use the exact lower bound for $x > 1$:
\begin{equation}
    T_k(x) \geq \frac{1}{2} \left(x + \sqrt{x^2 - 1}\right)^k\,.
\end{equation}
Substituting $x = 1 + 2\gamma_1$ yields
\begin{equation}
    T_{m-1}(1 + 2\gamma_1) \geq \frac{1}{2} \left( \sqrt{\gamma_1} + \sqrt{1 + \gamma_1} \right)^{2(m-1)}\,.
\end{equation}
For the large-scale physical Hamiltonians where iterative methods are most essential~\cite{Sandvik2010}, the relative gap is typically vanishingly small ($\gamma_1 \ll 1$). This occurs because the energy gap to the first excited state is usually microscopic compared to the extensive total width of the spectrum. In this regime, we obtain a highly accurate approximation by Taylor expanding $\ln(\sqrt{\gamma_1} + \sqrt{1 + \gamma_1}) \approx \sqrt{\gamma_1}$, directly leading to the following bound:
\begin{equation}
    \frac{1}{[T_{m-1}(1 + 2\gamma_1)]^2} \lesssim 4 e^{-4(m-1)\sqrt{\gamma_1}}\,.
\end{equation}

The convergence of higher excited states $\varepsilon_j$ follows a similar trajectory, though it is subject to a delay as the algorithm must first resolve the lower-lying spectrum. The error for the $j$-th eigenvalue is bounded by~\cite{Saad1980block}
\begin{equation}
    0 \leq \varepsilon_j - E_j \leq (E_{\mathcal{D}} - E_1) \mathcal{C}_j \frac{\tan^2 \theta_j}{[T_{m-j}(1 + 2\gamma_j)]^2}\,,
\end{equation}
where $\gamma_j = (E_{j+1} - E_j)/(E_{\mathcal{D}} - E_{j+1})$ is the relative gap to the next excited state. The prefactor $\mathcal{C}_j = \prod_{k=1}^{j-1} \left(\frac{E_{\mathcal{D}} - E_k}{E_j - E_k}\right)^2$ represents a convergence penalty that increases with $j$, and can be interpreted as the difficulty of separating the target from the $j-1$ lower lying states by the Lanczos algorithm. Furthermore, the Chebyshev term itself reflects a slower rate for interior states: the relative gap $\gamma_j$ is typically smaller than $\gamma_1$, and the reduced polynomial order $m-j$ effectively delays the convergence by $j$ Lanczos iterations compared to the ground state.

Let us note that the presented logic also applies to the reversed ordering of eigenvalues when the highest excited states are considered. For the approximately symmetric spectra found in typical many-body systems, this implies that the outer edges of the spectrum converge first, with the interior eigenvalues following after an increasing number of Lanczos steps $m$.

\subsubsection*{Convergence of Block Lanczos}
The standard Kaniel-Paige theory reveals a vulnerability: if the target state is exactly degenerate (e.g., $E_1 = E_2$), the relative gap $\gamma_1$ vanishes to zero. Furthermore, even without exact degeneracies, highly clustered extremal states (small $E_2 - E_1$) severely throttle the exponential convergence rate. The block Lanczos method~\cite{Underwood1975, Saad1980block} bypasses these limitations in exact arithmetic by expanding the Krylov subspace using a block of $s$ orthonormal starting vectors, while also allowing for the parallelization of calculations. The logic governing convergence is identical to the standard Lanczos algorithm, with the explicit error bound for any state $i$ within the target block ($1 \leq i \leq s$) taking the analogous form~\cite{Saad1980block},
\begin{equation}
    0 \leq \varepsilon_i - E_i \leq \frac{(E_\mathcal{D} - E_i)\tan^2 \theta_i}{[T_{m-1}(1 + 2\gamma_i^{(s)})]^2}\,,
\end{equation}
where $\theta_i$ is the principal angle between the exact eigenvector $\ket{u_i}$ corresponding to $E_i$ and the subspace spanned by the initial block of $s$ starting vectors. The generalized block relative gap is defined as:
\begin{equation}
    \gamma_i^{(s)} = \frac{E_{s+1} - E_i}{E_\mathcal{D} - E_{s+1}}\,.
\end{equation}
By pushing the boundary of the unwanted bulk spectrum up from $E_2$ to $E_{s+1}$, the algorithm becomes entirely immune to internal degeneracies of multiplicities up to $s$ in exact arithmetic. Moreover, typically $\gamma_i^{(s)} \geq \gamma_i$: by artificially inflating the gap, the block method accelerates the exponential decay of the error bound, significantly reducing the total iteration count $m$ required for convergence. However, because the computational cost per iteration of the block method is greater than that of the standard Lanczos algorithm, it usually yields similar or slightly lower overall computational efficiency for the same number of converged states, representing a fundamental trade-off between robustness, parallelizability, and raw performance.

The convergence behavior of higher excited states follows the same logic as established for the standard algorithm, with blocks of interior states experiencing delayed convergence relative to the ground-state block. Furthermore, this framework applies equally to the highest excited states; for approximately symmetric spectra, these upper extremal blocks converge at a rate comparable to the ground-state block.

\section{Methods for targeting the central part of the spectrum}
\label{Methods:for:targeting:central:part:of:the:spectrum}

To obtain eigenpairs from the middle of the spectrum without resorting to full exact diagonalization, several options are available: general iterative eigensolvers~\cite{Hernandez05slepc, Bollhofer07jadamilu, Saad11eigenvalue}, Krylov-type eigensolvers~\cite{Lanczos1950,Ericsson80spectral, Saad11eigenvalue, Pietracaprina2018, Sierant20polfed}, and filtered-subspace eigensolvers~\cite{Polizzi2009FEAST, Tang2014FEAST, Guan2021Dual}, the latter two being the most popular in the field of quantum many-body physics.
Obtaining eigenpairs from the middle of the spectrum with Krylov-type methods requires some form of spectral transformation. As discussed in the previous section, the Lanczos algorithm naturally converges to the extremal eigenvalues and the corresponding eigenstates of a system, so accessing different parts of the spectrum requires applying a spectral transformation to the Hamiltonian~\cite{Ericsson80spectral, Saad11eigenvalue}. Secs.~\ref{sec:microcanonical:lanczos:method} and~\ref{sec:polynomial:spectral:filtering} below present methods based on polynomial filtering, whereas Sec.~\ref{sec:shift:and:invert} presents a method based on rational filtering.
Another possibility is to employ a filtered-subspace algorithm, which projects out the irrelevant part of the spectrum and thereby exposes the desired one. Since this work focuses on Krylov-type methods, we provide details of the two most popular solutions of this kind in Appendix~\ref{app:feast:dacp}, and we provide brief summary of their applicability and limitations in Sec.~\ref{sec:comparison:of:the:methods}.

\subsection{Microcanonical Lanczos method} \label{sec:microcanonical:lanczos:method}
In the context of quantum many-body physics, the microcanonical Lanczos method \cite{PhysRevB.68.235106,avella2013} was developed for calculating the dynamical properties of systems at finite temperature. This method uses the transformation $(H-\lambda \mathds{1})^2$ to target energies around $\lambda$. Under the $(H-\lambda\mathds{1})^2$ spectral transformation, the spacing between neighboring eigenvalues near $\lambda$ remains exponentially small in the system size. Consequently, the number of iterations needed for convergence in the bulk of the spectrum is prohibitively large. This problem is mitigated with two compromises. First, the method does not store all previously computed Lanczos vectors, which is required for reorthogonalization. Instead, one runs the algorithm twice, as in the finite-temperature Lanczos method (FTLM) \cite{PhysRevB.finite-temperature-Lanczos,avella2013}. Second, one does not perform as many iterations as needed for full convergence of eigenvalues and eigenstates, accepting only rough convergence, which is sufficient for computing thermodynamic properties. As a result, this approach does not yield individually converged eigenstates. Nevertheless, one obtains a representative state from the targeted microcanonical window due to the effective averaging over nearby eigenstates.
    
\subsection{The shift-and-invert method} \label{sec:shift:and:invert}

Another numerical method, widely used in the many-body community, is the shift-and-invert method~\cite{Ericsson80spectral}, which provides a fast and stable way of calculating eigenpairs from the desired part of the spectrum.

The shift-and-invert method~\cite{Ericsson80spectral} constructs the transformed matrix $H^\prime = (H - \lambda \mathds{1})^{-1}$, whose eigenvalues $E^\prime = (E - \lambda)^{-1}$ are large for eigenvalues $E$ close to the target energy $\lambda$, while the eigenvectors remain unchanged. Consequently, states near the target become extremal eigenpairs of $H^\prime$, which strongly accelerates Krylov methods such as Lanczos or Arnoldi~\cite{Arnoldi51} algorithms. This is the main advantage of shift-and-invert with respect to the microcanonical Lanczos approach, which does not produce such a favorable spectral reordering. The method was applied to the diagonalization of many-body Hamiltonians, e.g., in Refs.~\cite{Luitz15, Serbyn17, Mace18, Pietracaprina2018, Suntajs20, Colbois23, Colbois24, Laflorencie25}.
    
The inversion is never evaluated explicitly; instead, each application of $H^\prime$ to a vector requires solving a linear system $(H - \lambda \mathds{1})\mathbf{x} = \mathbf{b}$. In practice, this is accomplished by computing a sparse LU decomposition of $(H - \lambda \mathds{1})$ using a direct solver such as MUMPS~\cite{Amestoy01mumps}. While LU decomposition preserves the exact eigenvalues, it introduces significant fill-in: the factors $L$ and $U$ are much denser than the original sparse Hamiltonian, leading to memory requirements that prohibit reaching larger system sizes.
 
An alternative is to use an incomplete LU (ILU) factorization as a preconditioner within iterative eigenvalue methods such as Jacobi--Davidson~\cite{Bollhofer07jadamilu}. This approach has been successfully applied to single-particle Anderson localization problems~\cite{Rodriguez10anderson, Rodriguez11multifractal}, where the Hamiltonian is a sparse tight-binding matrix with modest bandwidth. However, for interacting many-body Hamiltonians, the denser connectivity structure of the matrix has so far prevented ILU-based approaches from achieving comparable success.

\subsection{POLFED method} \label{sec:polynomial:spectral:filtering}
\begin{figure}[b!]
    \centering
    \includegraphics[width=0.99\textwidth]{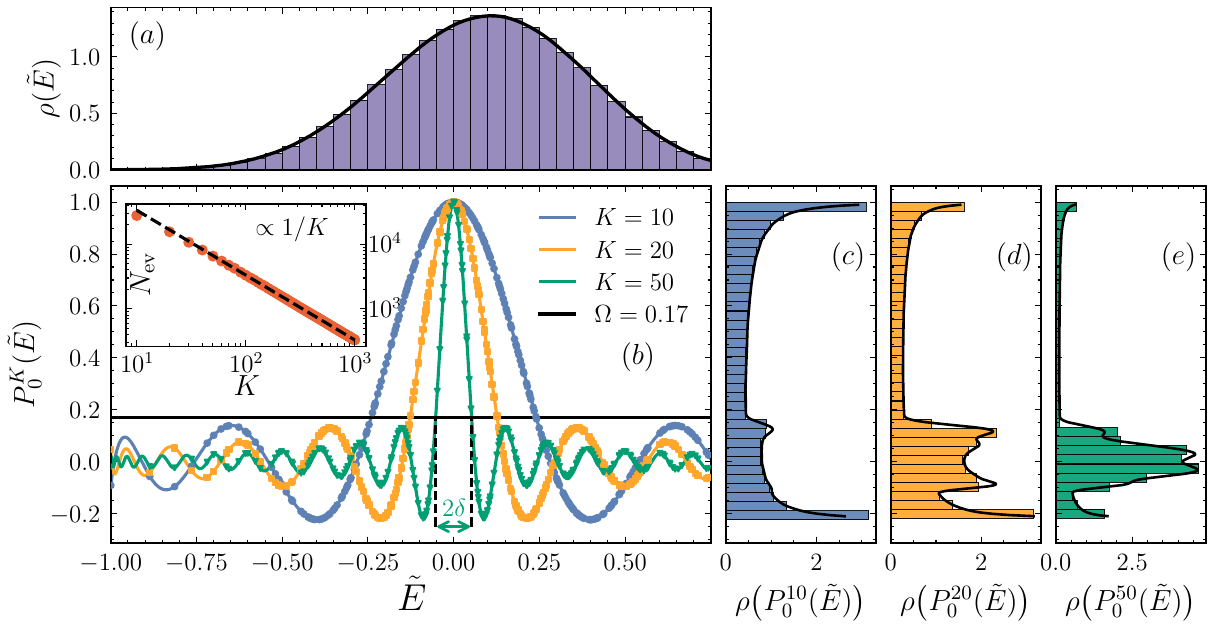}
    \caption{\textbf{Spectral transformation.} (a) Normalized density of states $\rho(\tilde{E})$ of the XXZ model, see Eq.~\eqref{eq:xxz}, with $L=18$ sites. (b) Spectral transformation $P_{\tilde\lambda}^{K}(\tilde{H})$, see Eq.~\eqref{eq:transformation}, for different orders of polynomial expansion $K=10, 20,$ and $50$ at target energy $\tilde\lambda=0$. The black horizontal line denotes the cutoff value $\Omega$ and $2\delta$ denotes the width of the energy window whose transformed eigenvalues lie above $\Omega$; both parameters play a key role in determining $K$, see Sec.~\ref{sec:POLFED}. The inset shows the number of eigenvalues $N_{\rm ev}$ as a function of $K$; for clarity, transformed eigenvalues in the main panel (b) are plotted for $L=10$ sites. (c,d,e) Density of states of the transformed XXZ Hamiltonian with $L=18$ sites in the zero-magnetization sector, for polynomial order $K=10,20,$ and $50$, respectively. Black solid lines in panels (a), (c), (d), and (e) show the approximate density of states obtained with the kernel polynomial method (KPM), as described in Sec.~\ref{sec:dos}.}
\label{fig:EigenvalueTransformation}
\end{figure}

To introduce polynomial spectral filtering of $H$, consider a matrix function $f$ of $H$ defined by a convergent power series,
\begin{align}
    \label{eq:taylor}
    f(H) = \sum_{n=0}^{\infty} a_n H^n.
\end{align}
Inserting the eigendecomposition, $H = P \, E \, P^{-1},$ one obtains
\begin{align}
    \label{eq:analytical_function_expension}
    f(H) = f(P E P^{-1}) = \sum_{n=0}^{\infty} a_n (P E P^{-1})^n = P \sum_{n=0}^{\infty} a_n E^n P^{-1} = P f(E) P^{-1}.
\end{align}
Equation~\eqref{eq:analytical_function_expension} implies that the eigenvectors of the transformed and original Hamiltonian coincide, while the eigenvalues are mapped as $E_i \mapsto f(E_i)$. Compared to shift-and-invert, a polynomial spectral transformation preserves the sparsity of individual matrix-vector products, as we discuss below.
        
In POLFED, we use a spectral transform based on the expansion of the Dirac delta function in the Chebyshev polynomials. The Chebyshev polynomials of the first kind are defined by the recurrence relation
\begin{align}
    \label{eq:chebyshev_definition}
    T_0(\tilde{H}) &= \mathds{1}, \\
    T_1(\tilde{H}) &= \tilde{H}, \\
    T_{n+1}(\tilde{H}) &= 2 \tilde{H}\,T_n(\tilde{H}) - T_{n-1}(\tilde{H}).
\end{align}
Since the Chebyshev polynomials are defined on $[-1,1]$, the Hamiltonian must first be rescaled so that its spectrum lies within this interval:
\begin{align}
    \label{eq:rescaledMatrix}
    \tilde{H} = \frac{H-\mathds{1}E_c}{\Delta E},\qquad E_c=\frac{E_{\max}+E_{\min}}{2},\qquad \Delta E=\frac{E_{\max}-E_{\min}}{2}
\end{align}
From here on, all quantities denoted with $\tilde{\bullet}$ are assumed to be normalized by the same procedure \eqref{eq:rescaledMatrix}. Note that the extremal energies $E_{\mathrm{min}}$ and $E_{\mathrm{max}}$ can be efficiently computed using the standard Lanczos method, which converges rapidly for eigenpairs at the spectral extrema.
 
The Chebyshev expansion of the spectral filter takes the form
\begin{align}
    \label{eq:transformation}
    P_{\tilde{\lambda}}^{K}(\tilde{H}) = \frac{1}{\chi} \sum_{n=0}^{K} c_n^{\tilde{\lambda}} T_n(\tilde{H}).
\end{align}
Here, $\lambda$ ($\tilde{\lambda}$) is the target energy (rescaled target energy), $K$ is the order of the polynomial expansion, and $\chi$ ensures normalization, such that $P_{\tilde{\lambda}}^K(\tilde{\lambda})=1$. The coefficients $c_n^{\tilde{\lambda}}$ correspond to the Chebyshev expansion of the Dirac delta function $\delta(\tilde{H} - \tilde{\lambda}\mathds{1})$ and are given by
\begin{align}
    \label{eq:transformationCoefficients}
    c_n^{\tilde{\lambda}} = \sqrt{4-3\delta_{0,n}}\cos(n\arccos(\tilde{\lambda}))=\sqrt{4-3\delta_{0,n}}T_n(\tilde\lambda).
\end{align}

The major computational bottlenecks of the spectral transformation
are related to the polynomial degree $K$ and the loss of sparsity for a naive implementation. In Fig.~\ref{fig:EigenvalueTransformation}(b), we show how the spectrum of the rescaled Hamiltonian transforms for different polynomial orders $K$. Since the filter approximates a Dirac delta function, the spectral weight of $P_{\tilde \lambda}^K(\tilde{H})$ becomes increasingly concentrated near the target energy $\tilde \lambda$ with increasing $K.$ In a typical setting, we adjust $K$ so that a fixed number $N_\mathrm{ev}$ of eigenpairs in the transformed spectrum lies above a chosen cutoff threshold $\Omega.$ We set $\Omega$ slightly above the height of the secondary peaks of $P_{\tilde \lambda}^K(\tilde{E})$, as indicated by the black horizontal line in Fig.~\ref{fig:EigenvalueTransformation}(b). Due to the exponential growth of the quantum many-body density of states with the system size $L,$ the required $K$ for fixed $N_\mathrm{ev}$ grows exponentially with $L$. The width $\delta$ of the energy window $[\tilde\lambda-\delta, \tilde\lambda+\delta]$, in which transformed energies lie above $\Omega$, scales inversely with the polynomial order, $\delta \propto 1/K$, and linearly with the number of requested eigenpairs $N_{\rm{ev}}$. In the inset of Fig.~\ref{fig:EigenvalueTransformation}(b), one can see that number of requested eigenpairs is inversely proportional to the order of polynomial expansion $N_{\rm{ev}} \propto 1/K$. Since the many-body density of states grows exponentially with $L$, maintaining a fixed $N_\mathrm{ev}$ above $\Omega$ requires $K$ growing with the Hilbert-space dimension.

\begin{figure}[t!]
    \centering
    \includegraphics[width=0.999\textwidth]{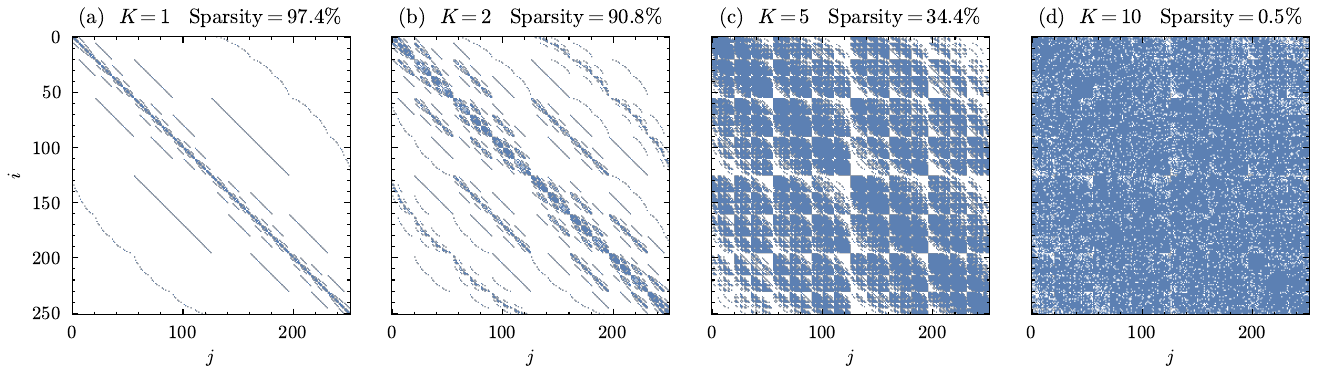}
    \caption{\textbf{Sparsity of the untransformed and transformed Hamiltonian.} Sparsity pattern of $P_{\tilde\lambda}^K(\tilde{H})$ for different polynomial orders $K$, where sparsity is measured as the fraction of zero matrix elements. Panels (a), (b), (c), and (d) correspond to polynomial orders $K=1,2,5,$ and $10$, respectively. For $K=1$ the filter is linear in $\tilde{H}$, so the sparsity pattern coincides with that of the original Hamiltonian. Sparsity is rapidly lost with each additional order of polynomial expansion. Results are shown for the XXZ model in the zero-magnetization sector with $L=10$ sites and target energy $\tilde\lambda=0$.}
    \label{fig:HamiltonianTransformation}
\end{figure}

Figs.~\ref{fig:EigenvalueTransformation}(c)-\ref{fig:EigenvalueTransformation}(e) show how the density of states of the transformed Hamiltonian changes with increasing $K$. The peak in the density of states at $P^K_0(\tilde{E}) = 1$ is significantly reduced with increasing $K$, resulting in fewer iterations needed for the Lanczos algorithm to resolve nearby eigenvalues---i.e., faster convergence. This peak originates from the smooth behavior of the polynomial transformation near $\tilde{\lambda}$ (continuous second derivative) and cannot be avoided entirely. Thanks to its reduction with $K$, finding the degree $K$ such that the interval $[\tilde \lambda-\delta, \tilde\lambda+\delta]$ contains exactly $N_{\rm ev}$ eigenvalues ensures optimal convergence. Between the peak at $P^K_0(\tilde{E}) = 1$ and the cutoff $\Omega$, the density of states is nearly flat. This ensures large relative spectral gaps $\gamma_j$, which, according to the Kaniel-Paige bounds (Sec.~\ref{sec:convergence:krylov:methods}), guarantee rapid convergence for eigenpairs in this target window. A second, more pronounced peak appears slightly below $\Omega$, corresponding to the secondary maximum of the polynomial filter. Crucially, the choice of a Chebyshev expansion guarantees that the transformed spectrum approaches this ideal uniformity as fast as possible with increasing polynomial degree $K$, a direct consequence of the optimal approximation properties of Chebyshev polynomials over the bounded spectral interval.

A second challenge is the \emph{loss of sparsity}, as illustrated in Fig.~\ref{fig:HamiltonianTransformation}. The transformed matrix becomes fully dense at relatively small polynomial orders, $K\sim 10$, whereas in typical calculations one needs $K\sim 1000$ or more. An explicit implementation of the spectral transform in terms of repeated matrix-matrix multiplications would yield a dense transformed matrix, with computational (and memory) costs much worse than with the shift-and-invert approach. Instead, in POLFED, the spectral transform is performed \emph{on the fly} at each step of the block Lanczos iteration. The multiplication of vectors by $P_{\tilde\lambda}^K(\tilde{H})$ is then decomposed into a sequence of $K$ sparse matrix-vector products with $\tilde{H}$, preserving the sparsity of the original Hamiltonian throughout. In practice, the polynomial sum is evaluated in a numerically stable manner using the Clenshaw algorithm, which we introduce in Sec.~\ref{sec:clenshaw}.

Performing exponentially many (in system size) matrix products could lead to a loss of accuracy in finite-precision arithmetic. We show in Appendix~\ref{app:accuracy:and:reliability} that performing the mapping with the Clenshaw algorithm is numerically stable and reliable.

\subsection{Comparison of the methods} \label{sec:comparison:of:the:methods}

In this section we provide qualitative comparison of the methods, with quantitative benchmarks discussed in Sec.~\ref{sec:benchmarks}. Firstly, we compare two Krylov-type methods: shift-and-invert and POLFED, noting that microcanonical Lanczos is now obsolete for most applications due to its low precision. The main drawback of shift-and-invert is the very high memory consumption of the LU factorization, which exceeds the memory available on a single computing node already for modest system sizes, $L\approx 20$ (without resolving symmetries, the Hilbert-space dimension is $\mathcal{D}=2^{20}$). Both the memory requirements and the runtime grow rapidly with the number of off-diagonals present in the matrix. Consequently, multi-node (MPI) parallelization is required, which adds complexity to code implementation, requires substantial high-performance-computing expertise, and complicates job submission. Pushing to larger system sizes, even well-equipped HPC facilities may be unable to store the factorization: for the XXZ model at $L=26$, $204$~TB of memory were required, see Ref.~\cite{Pietracaprina2018}. On the upside, the algorithm is highly parallelizable and scales well across multiple nodes, so that it can make use of up to tens of thousands of CPU cores. Moreover, shift-and-invert can be used for diagonalization of general matrices, while POLFED is only applicable to systems with eigenvalues occupying a one-dimensional manifold of the complex plane, most notably hermitian and unitary matrices.

We now compare POLFED with the filtered-subspace algorithms, discussed in more detail in Appendix~\ref{app:feast:dacp}. Starting with FEAST~\cite{Polizzi2009FEAST,Tang2014FEAST}, see Appendix~\ref{sec:feast:method}, the method suffers from rapidly growing memory requirements, which makes it impractical for single-node computations. It does not offer a shorter time to solution either, but its advantages are twofold: good parallel scaling (as for shift-and-invert) and, most importantly, insensitivity to degeneracies, which pose a notable problem for Krylov methods. The method is also applicable to general diagonalizable matrices, unlike POLFED. The second filtered-subspace algorithm is DACP~\cite{Guan2021Dual}, see Appendix~\ref{sec:dacp}, which is fast and memory efficient, but lacks the accuracy required for resolving individual eigenvalues.

Finally, POLFED is characterized with a low memory footprint, fast execution, high accuracy and a simple workflow that does not require multi-node scaling. Its shortcomings are the less favorable parallel scaling to large numbers of CPU cores and the large polynomial orders required for large system sizes, as well as limited applicability to hermitian and unitary matrices.

Implementations of several of the methods discussed above are publicly available. Among these, \inh{Polfed.jl} is characterized by straightforward installation process and ability to treat fairly large system sizes on a modest laptop. Moreover, \inh{Polfed.jl} provides GPU support, which, to the best of our knowledge, no other publicly available library targeting interior eigenpairs of sparse Hamiltonians currently offers.

\section{POLFED algorithm}
\label{sec:POLFED}

In this section, we describe the individual steps of the POLFED algorithm. We first explain how the density of states is estimated (Sec.~\ref{sec:dos}) and how it determines the polynomial order $K$ (Sec.~\ref{sec:order:of:spectral:filtering}). We then detail the block Lanczos iteration with the polynomial filter applied on the fly (Sec.~\ref{sec:blocklanczosalgorithm}), including the Clenshaw recurrence used for its numerically stable evaluation (Sec.~\ref{sec:clenshaw}).

\subsection{Finding density of states}
\label{sec:dos}
To compute $N_{\rm{ev}}$ eigenpairs, the POLFED algorithm requires an estimate of the density of states (DOS) to determine the polynomial order $K$. Computing the DOS exactly would require full diagonalization, which defeats the purpose. However, several approximate methods~\cite{densetiesofstates, Silver96} can estimate the density of states without full diagonalization. The simplest and often adequate approach for interacting many-body systems is to assume a Gaussian density of states:
    \begin{align}
    \label{eq:dos_gaussian}
        \rho(\tilde{E}) = \frac{\mathcal{D}}{\sqrt{2\pi}\tilde{\Gamma}} \exp\left[-\frac{(\tilde{E} - \tilde{\mu})^2}{2\tilde{\Gamma}^2}\right],
    \end{align}
where $\tilde{E}$ is the rescaled energy, and $\tilde{\mu}$ and $\tilde{\Gamma}$ denote the mean and standard deviation of the spectrum of $\tilde{H}$, respectively. 

At a higher computational cost, a more accurate estimate can be obtained using the kernel polynomial method (KPM)~\cite{kpm, densetiesofstates, Silver96}, whose cost scales linearly with the Hilbert-space dimension $\mathcal{D}$ for a given energy resolution. The method expands an arbitrary function $f$ in Chebyshev polynomials with coefficients $\mu_n$ (also called moments), and suppresses Gibbs oscillations by multiplying each moment with a damping factor $g_n$:
    \begin{align}
        f_{\rm KPM}(x) = \frac{1}{\pi \sqrt{1-x^2}}\bigg[ \mu_0 g_0 + 2 \sum_{n=1}^{N-1} \mu_n g_n T_n(x) \bigg]\,, \quad \quad \quad \mu_n = \int_{-1}^{1} f(x) T_n(x) \text{d}x.
    \end{align} 
We provide further details in Appendix~\ref{app:kpm}; for a comprehensive review, see Ref.~\cite{kpm}.
 
\begin{figure}[htb!]
        \centering
        \includegraphics[width=0.75\textwidth]{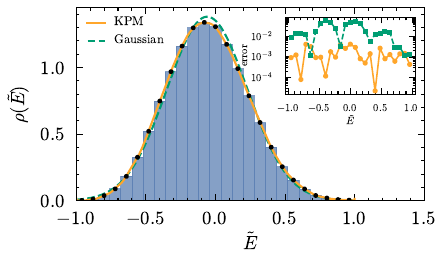}
    \caption{\textbf{Density of states}. We consider the $J_1$--$J_2$ model, see Eq.~\eqref{eq:j1j2}, for $L=20$ in the magnetization sector with $N_\uparrow=5$ up spins ($S^z_{\rm tot}=-5$). The blue histogram shows results obtained from exact diagonalization (ED); black dots at the bin centers are shown to facilitate comparison with the KPM and Gaussian approximations. The KPM and Gaussian curves are computed from Eqs.~\eqref{eq:dos_kpm} and \eqref{eq:dos_gaussian}, and are shown with orange and green lines, respectively. The KPM estimate uses $R=100$ random vectors and $75$ Chebyshev moments. The inset shows the absolute error of both approximations at the bin centers, evaluated against ED. Note that the quantitative values of the errors slightly depend on the binning of the histogram.}
        \label{fig:dos_comparisment}
\end{figure}

The density of states $\rho(\tilde{E}) = \sum_{k=1}^{\mathcal{D}} \delta(\tilde{E} - \tilde{E}_k)$ can be approximated with KPM as
    \begin{align}
        \label{eq:dos_kpm}
        \rho_{\rm KPM}(\tilde{E}) = \frac{1}{\pi \sqrt{1-\tilde{E}^2}}\bigg[ \mu_0 g_0 + 2 \sum_{n=1}^{N-1} \mu_n g_n T_n(\tilde{E}) \bigg]\,,
    \end{align}
    with moments: 
    \begin{align}
        \mu_n = \sum_{k=1}^{\mathcal{D}} \int_{-1}^{1} \delta(\tilde{E} - \tilde{E}_k) T_n(\tilde{E}) \text{d}\tilde{E} =  \sum_{k=1}^{\mathcal{D}} T_n(\tilde{E}_k) =  \sum_{k=1}^{\mathcal{D}} \bra{u_k} T_n(\tilde{H}) \ket{u_k} = \Tr \big\{ T_n(\tilde{H})\big\},
    \end{align}
    where $\ket{u_k}$ represents the exact eigenstates of $H$. The $n$-th moment of the density of states is thus equal to the trace of the $n$-th Chebyshev polynomial of the Hamiltonian. In \inh{Polfed.jl}, these traces are estimated stochastically as 
\begin{equation}
        \Tr\{ T_n(\tilde{H})\}= \frac{1}{R}\sum_{r=1}^R \bra{r} T_n(\tilde{H})\ket{r},
\end{equation}
using $R$ random vectors $\ket{r}$ whose components are independent random variables with zero mean and unit variance. For the kernel function $g_n$, we use the Jackson kernel:
    \begin{align}
        g_n = \frac{1}{N+1}\bigg[(N-n+1)\cos(\frac{\pi n}{N+1}) + \sin(\frac{\pi n}{N+1})\cot(\frac{\pi}{N+1})  \bigg],
    \end{align}
    although the Lanczos kernel \eqref{eq:kernels_jackson_and_lorenz} works adequately as well. Figure~\ref{fig:dos_comparisment} shows a comparison between the Gaussian approximation, the KPM, and the density of states obtained from exact diagonalization for the $J_1$--$J_2$ model. The KPM approximation is considerably more accurate than the Gaussian one. Furthermore, the Gaussian approximation fails to describe asymmetric profiles by construction. For this reason, \inh{Polfed.jl} uses the KPM estimate by default.

\subsection{Order of polynomial expansion}
\label{sec:order:of:spectral:filtering}
The polynomial order $K$ controls the number of eigenvalues $N_{\rm{ev}}$ that lie in the interval $[\tilde{\lambda}-\delta, \tilde{\lambda}+\delta]$ above the cutoff $\Omega$, as illustrated in Fig.~\ref{fig:EigenvalueTransformation}. To determine $K$ for a given $N_\mathrm{ev}$, one needs the density of states $\rho(\tilde{E})$ near the target energy. For large systems, the energy window $2\delta$ becomes exponentially small, and one can make the microcanonical approximation: 
\begin{align}
    \label{eq:Nev}
    N_{\rm{ev}} = \int_{\tilde{\lambda}-\delta}^{\tilde{\lambda}+\delta} \rho(\tilde{E}) \text{d}\tilde{E}  \approx 2\delta \rho(\tilde{\lambda}) \qquad \rightarrow \qquad \delta = \frac{N_{\rm{ev}}}{2\rho(\tilde{\lambda})}.
\end{align}

Although the microcanonical approximation in Eq.~\eqref{eq:Nev} is already accurate for large Hilbert spaces, \inh{Polfed.jl} does not rely on it and instead evaluates the integral in closed form. No numerical quadrature is required: the KPM estimate~\eqref{eq:dos_kpm} is a Chebyshev series, and with the substitution $\tilde{E}=\cos\theta$, each term can be integrated analytically,
\begin{align}
    \label{eq:chebyshev:int}
    \int \frac{T_n(\tilde{E})}{\pi\sqrt{1-\tilde{E}^{2}}}\,\mathrm{d}\tilde{E} =
    \begin{cases}
        -\dfrac{1}{\pi}\arccos(\tilde{E}), & n=0,\\[8pt]
        -\dfrac{1}{n\pi}\sin\!\big(n\arccos(\tilde{E})\big), & n\geq 1.
    \end{cases}
\end{align}
The number of eigenvalues $N_{\rm{ev}}$ inside the target window therefore follows directly from the moments $\mu_n$ and the kernel coefficients $g_n$,
\begin{align}
    \label{eq:Nev:exact}
    N_{\rm{ev}} = \frac{1}{\pi}\bigg[ \mu_0 g_0 \big(\theta_- - \theta_+\big)
    + 2\sum_{n=1}^{N-1} \frac{\mu_n g_n}{n} \big(\sin n\theta_- - \sin n\theta_+\big) \bigg],
    \qquad \theta_\pm = \arccos(\tilde{\lambda}\pm\delta),
\end{align}
which is solved for $\delta$ with bisection. 
We note that alternative approaches, e.g., quadrature rule such as Simpson's rule, may also be used to obtain the number of eigenvalues $N_{\rm{ev}}$.
The advantage of the method described above is to provide the closed-form expression for $N_{\rm{ev}}$, and it introduces no discretization error.

The polynomial order $K$ in POLFED is then fixed by requiring that the transformed eigenvalues at the edges of the window equal the cutoff value $\Omega$,
    \begin{align}
        \label{eq:condition:for:K}
        P^K_{\tilde{\lambda}}\left(\tilde{\lambda} \pm \delta\right) = \Omega.
    \end{align}    
Equation~\eqref{eq:condition:for:K} can be solved for $K$ using a bisection search between $K_{\mathrm{min}} = 1$ and $K_{\mathrm{max}} = \mathcal{D}$, or by incrementing $K$ until Eq.~\eqref{eq:condition:for:K} is satisfied. The scaling of $K$ can also be estimated semi-analytically. Since the filter width scales as $\delta \propto 1/K$, and $\delta$ itself depends approximately on $\rho(\tilde{\lambda})$ through Eq.~\eqref{eq:Nev}, the dependence of $K$ on $\tilde{\lambda}$ takes the form $K(\tilde{\lambda}) = f(\tilde{\lambda})/\delta(\tilde{\lambda}, N_{\rm ev})$, where $f(\tilde{\lambda})$ is an additional factor arising from the properties of the Chebyshev expansion at $\tilde{\lambda}$.
    
Empirically, we find that $f(\tilde{\lambda})$ is well approximated by a semicircle, leading to the expression:
    \begin{align}
    \label{eq:K:approx}
    K(\tilde{\lambda}) = \frac{2.66}{\delta(\tilde{\lambda},N_{\rm ev})} \sqrt{1 - \tilde{\lambda}^2}.
    \end{align}
Since the semicircle dependence is obtained empirically from tests on specific Hamiltonians, Eq.~\eqref{eq:K:approx} should be understood as a practical approximation rather than a universal relation. We show the accuracy of the empirical Eq.~\eqref{eq:K:approx} in Appendix~\ref{app:details:on:the:polynomial:order}.

\subsection{Block Lanczos algorithm}
\label{sec:blocklanczosalgorithm}

\begin{figure}[b!]
    \centering
    \includegraphics[width=0.85\textwidth]{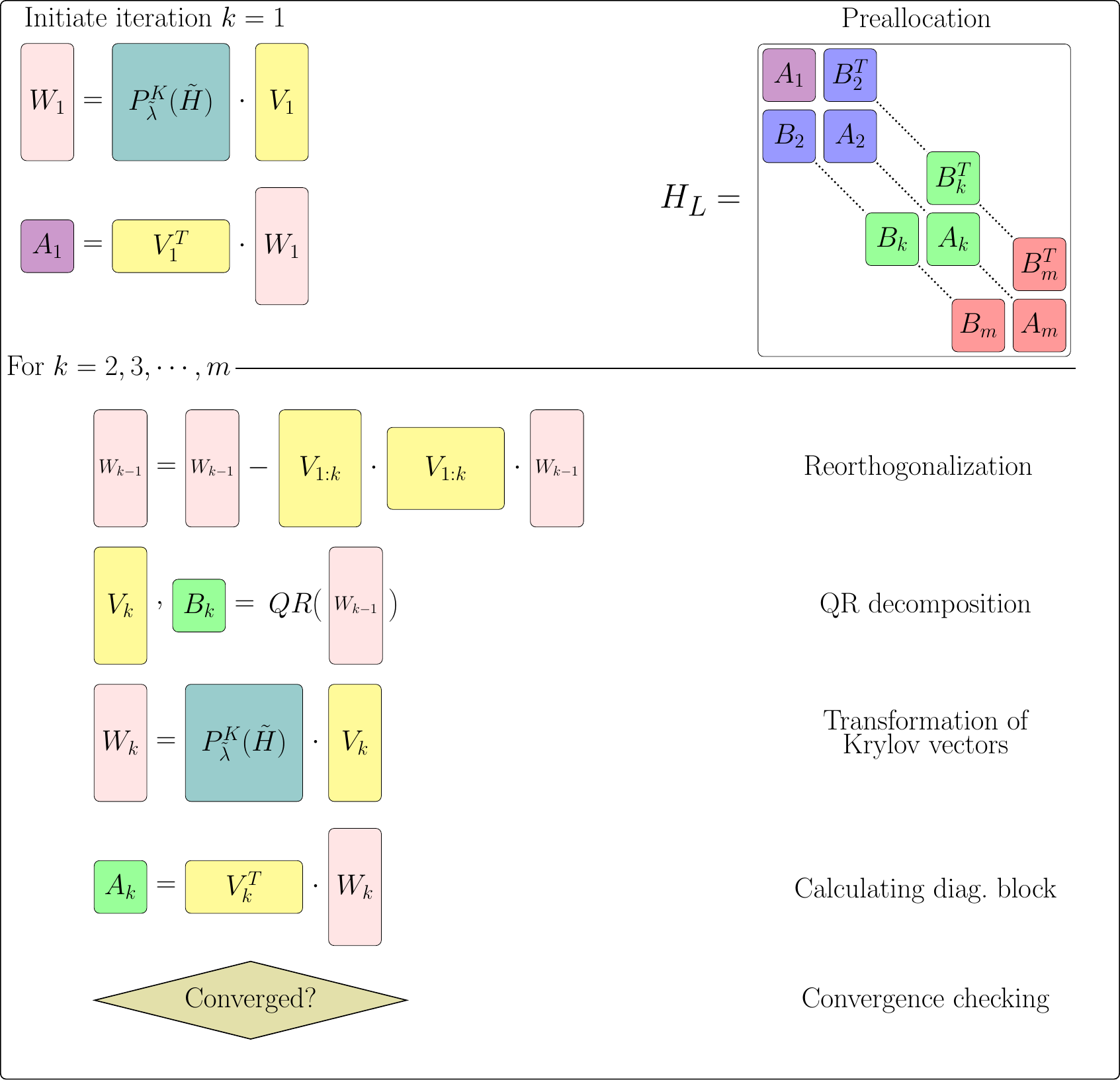}    
    \caption{\textbf{Schematic overview of the block Lanczos algorithm.} The inputs are the initial Krylov block $V_1$ and the polynomial spectral filter $P_{\tilde{\lambda}}^K(\tilde{H})$.}
    \label{fig:BlockLanczos}
\end{figure}
    
We now describe the block Lanczos iteration as implemented in \inh{Polfed.jl}. Figure~\ref{fig:BlockLanczos} shows a pictorial representation along with the pseudocode. For optimal performance, it is advantageous to preallocate all required matrices, in particular the Lanczos matrix $H_L,$ which is a block-tridiagonal matrix of size $M \times M$ (where $M=m\cdot s$, $m$ is the number of block iterations and $s$ is the block size). Our implementation is written in the \href{https://julialang.org}{Julia} programming language, so the schematic in Fig.~\ref{fig:BlockLanczos} uses 1-based array indexing.
 
At $k=1$, we start with a block of $s$ orthonormal initial vectors $V_1 \in \mathbb{C}^{\mathcal{D} \times s}.$ Applying the polynomial filter yields the first transformed Krylov block, $W_1 = P_{\tilde{\lambda}}^K(\tilde{H})V_1$. The first diagonal block of $H_L$ is then obtained by projection, $A_1 = V_1^\dagger W_1$.
 
For the remaining steps, we proceed as follows. First, reorthogonalization of the transformed block $W_{k-1} = P_{\tilde{\lambda}}^K(\tilde{H}) V_{k-1}$ against all preceding Krylov blocks is performed. A QR decomposition of the orthogonalized $W_{k-1}$ then yields a new Krylov block $V_k$ and the off-diagonal block $B_k$ of the Lanczos matrix. The filter is then applied again, $W_k = P_{\tilde{\lambda}}^K(\tilde{H}) V_k$, and the diagonal block $A_k$ is computed by projection. Crucially, the spectral filter $P_{\tilde{\lambda}}^K(\tilde{H})$ is never explicitly formed as a matrix. Instead, only the product $P_{\tilde{\lambda}}^K(\tilde{H}) V_i$ is evaluated on the fly, preserving the sparsity of $H$. This is done in a numerically stable way using the Clenshaw algorithm, as mentioned in Sec.~\ref{Methods:for:targeting:central:part:of:the:spectrum} and discussed later in Sec.~\ref{sec:clenshaw}---this is the key feature of POLFED that avoids the memory overhead of shift-and-invert.
    
At the end of iteration, we diagonalize $H_L$ (obtaining eigenvectors $t_i$) to check whether the convergence criterion is met. Specifically, we require $\max_{i} (\norm{B_{k+1} \tau_{ki}})<\epsilon$, where $\tau_{ki}$ denotes the $k$-th block of size $s$ of $t_i$, and $\epsilon$ is the desired accuracy. In practice, convergence checking is not performed at every step but at intervals determined by the residuals from previous steps.

Once the eigenvectors of the Lanczos matrix are computed and the convergence criterion is fulfilled, we transform these eigenvectors back to the Hilbert-space basis via
    \begin{align}
    u_i = V_{1:m} t_i,
    \end{align}
where $V_{1:m}=[V_1,V_2,\ldots,V_m]$ denotes the horizontal concatenation of the first $m$ Krylov blocks. Here, $u_i$ are the eigenvectors\footnote{In the preceding sections, we rigorously distinguished between the approximate Ritz vectors $\psi_i$ and the exact Hamiltonian eigenvectors $u_i$. For the remainder of the text we will drop this distinction and use $u_i$ to denote the converged Ritz vectors.} of $P_{\tilde{\lambda}}^K(\tilde{H})$, and consequently of $H$, since polynomial transformations preserve eigenvectors, cf.\ Eq.~\eqref{eq:analytical_function_expension}. The corresponding eigenvalues are obtained from the Rayleigh quotient, $E_i = \bra{u_i} H \ket{u_i}$. Finally, the residual norms $\|H u_i - E_i u_i\|$ are computed to verify the convergence of each eigenpair.

\subsection{On-the-fly spectral transformation with Clenshaw algorithm}
\label{sec:clenshaw}
As discussed in Sec.~\ref{sec:polynomial:spectral:filtering}, explicitly forming $P^K_{\tilde{\lambda}}(\tilde{H})$ would destroy the sparsity of the Hamiltonian, see Fig.~\ref{fig:HamiltonianTransformation}. POLFED therefore evaluates the product $P^K_{\tilde{\lambda}}(\tilde{H}) \cdot V_k$ on the fly at each block Lanczos step, using only repeated sparse matrix-vector multiplications with $\tilde{H}$. The key ingredient for this evaluation is the Clenshaw algorithm, which we describe next.

\subsubsection*{Clenshaw algorithm}

The Clenshaw algorithm~\cite{Clenshaw55} evaluates a finite sum of functions satisfying a three-term recurrence, such as a Chebyshev series, without explicitly computing the individual basis functions.
 
Consider the weighted sum:
    \begin{align}
        S(x) =\sum_{k=0}^{K} a_k \phi_k(x)\,,
    \end{align}
where $\phi_k$ satisfy the recurrence $\phi_{k+1}(x) = \alpha_k(x)\,\phi_k(x) + \beta_k(x)\,\phi_{k-1}(x)$,
with known coefficients $\alpha_k(x)$ and $\beta_k(x)$. Rather than evaluating $\phi_k$ directly, one introduces auxiliary quantities $b_k(x)$ via the backward recurrence,
    \begin{align}
       b_{K+1}(x)&=b_{K+2}(x)=0,\\
       b_{k}(x)&=a_{k}+\alpha _{k}(x)\,b_{{k+1}}(x)+\beta _{{k+1}}(x)\,b_{{k+2}}(x).
    \end{align}
After computing $b_{2}(x)$ and $b_{1}(x)$, the sum is obtained from
    \begin{align}
        S(x)=\phi _{0}(x)\,a_{0}+\phi _{1}(x)\,b_{1}(x)+\beta _{1}(x)\,\phi _{0}(x)\,b_{2}(x).
    \end{align}

\subsubsection*{Implementation of Clenshaw algorithm}
\begin{figure}[t!]
    \centering
    \includegraphics[width=0.85\textwidth]{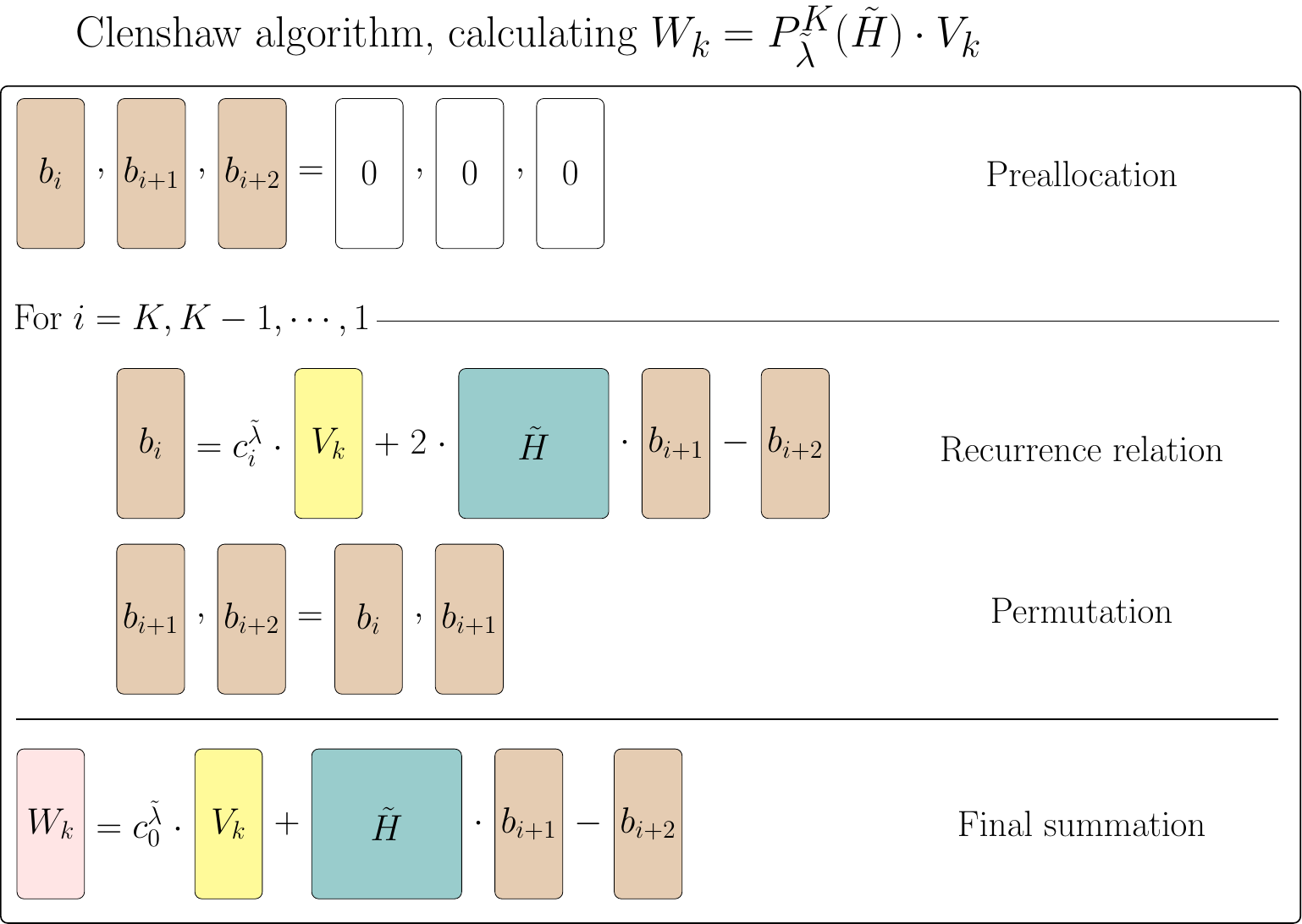}
    \caption{\textbf{Schematic of the Clenshaw recurrence algorithm.} This algorithm is used in POLFED to evaluate the polynomial filter on the fly. The inputs are the current Krylov block $V_k$ and the rescaled Hamiltonian $\tilde{H}$; the output is the transformed block $W_k = P_{\tilde{\lambda}}^K(\tilde{H})\, V_k$.}
\label{fig:ClenshawCartoon}
\end{figure}

In our case, the basis functions $\phi_k$ are the Chebyshev polynomials $T_k$, with recurrence coefficients 
    \begin{align}
        \alpha_k(x)=2x, \quad \quad \quad  \beta_k(x)=-1\,,
    \end{align}
and initial conditions 
\begin{equation}
T_0(x)=1, \qquad T_1(x)=x\,,
\end{equation}
cf.\ Eq.~\eqref{eq:chebyshev_definition}. The Clenshaw recurrence then simplifies to:
    \begin{align}
    \label{eq:clenshaw_chebyshev}
        b_{K+1}(x)&=b_{K+2}(x)=0 \\
        b_k(x) &= a_k + 2xb_{k+1}(x) - b_{k+2}(x) \notag \\
        S(x) &= a_0 + xb_1(x) - b_2(x). \notag
    \end{align}
To apply Eq.~\eqref{eq:clenshaw_chebyshev} to the matrix polynomial $P_{\tilde{\lambda}}^K(\tilde{H})$, one replaces $x$ by $\tilde{H}$. Since we only need the product $P_{\tilde{\lambda}}^K(\tilde{H})\, V_k$ (not the full matrix), the auxiliary quantities $b_i$ are $\mathcal{D} \times s$ blocks rather than $\mathcal{D} \times \mathcal{D}$ matrices, and scalar coefficients $a_i$ are multiplied by $V_k$. The recurrence becomes:
    \begin{align}
    \label{eq:clenshaw_chebyshev_prod}
        b_{K+1}&=b_{K+2}=0 \notag\\
        b_i &= a_i V_k + 2\tilde{H}\,b_{i+1} - b_{i+2} \\
        P_{\tilde{\lambda}}^K(\tilde{H}) V_k &= a_0 V_k + \tilde{H} b_1 - b_2. \notag
    \end{align}
This formulation requires only $K$ sparse matrix-block multiplications with $\tilde{H}$ and a storage of two auxiliary blocks, making it both time- and memory-efficient. A schematic of the procedure is shown in Fig.~\ref{fig:ClenshawCartoon}.

\subsection{Applications to GPUs}
\label{sec:gpu:polfed}
A natural application of POLFED is its implementation on graphics processing units (GPUs). This is advantageous for several reasons. First, unlike shift-and-invert methods, POLFED does not require storing or factorizing a transformed matrix. Instead, the filtering is performed on the fly through the polynomial recurrence discussed above. As a result, the method has a modest memory footprint. In practice, this often makes it possible to keep the full working set of the algorithm on a single GPU node, a regime that is much harder to reach with shift-and-invert. 

Second, the dominant computational cost of POLFED is the spectral transformation itself. In the on-the-fly formulation, applying the polynomial filter reduces to a sequence of repeated Hamiltonian applications to one or several vectors. The core workload consists of sparse matrix-vector and sparse matrix-matrix multiplications, which are well suited for GPU architectures. For the system sizes of interest, the performance of these operations is often limited not by floating-point throughput, but by memory bandwidth. Modern GPUs are particularly effective in this regime because they combine massive thread-level parallelism with high memory bandwidth.

The modest memory requirements of the method allow the data to remain on the device, while the dominant bottleneck, sparse-matrix operations, maps naturally onto GPU hardware. Consequently, the runtime can be substantially reduced even with a straightforward GPU implementation of the polynomial recurrence.

An additional advantage is that the spectral transformation can be optimized further when the structure of the Hamiltonian is known. Besides using a generic sparse-matrix representation, one may construct a mapping specialized for a given model, exploiting its particular connectivity or algebraic structure. Such specialized mappings avoid part of the overhead associated with general-purpose sparse linear algebra and can outperform standard sparse-matrix implementations. This is particularly useful in the polynomial-filtering setting, where the same mapping is applied repeatedly, and even modest improvements in a single application accumulate into a significant overall speedup. \inh{Polfed.jl} supports both generic sparse-matrix input and user-defined mappings, as described in Sec.~\ref{sec:interface}.

To summarize, GPUs are a natural platform for POLFED because the method combines a low memory footprint with a computational bottleneck, dominated by repeated sparse operations. This stands in contrast to shift-and-invert approaches, whose larger memory requirements make GPU implementations considerably more challenging.

\subsection{Summary of the POLFED workflow}
\label{sec:polfed:summary}

We conclude this section by summarizing the complete POLFED workflow. Given a sparse Hamiltonian $H$, a target energy $\lambda$, and a desired number of eigenpairs $N_\mathrm{ev}$, the algorithm proceeds as follows.
\begin{enumerate}
   \item \textbf{Spectral bounds.} Compute the extremal eigenvalues $E_\mathrm{min}$ and $E_\mathrm{max}$ using a standard Lanczos iteration and rescale the Hamiltonian to $\tilde{H}$ via Eq.~\eqref{eq:rescaledMatrix}.
    \item \textbf{Density of states.} Estimate the density of states $\rho(\tilde{E})$ near the target energy with the kernel polynomial method~\eqref{eq:dos_kpm}.
    \item \textbf{Half-width of the targeted interval.} With $\rho(\tilde{E})$ already calculated, find the half-width $\delta$ of the interval $[\tilde{\lambda}-\delta, \tilde{\lambda}+\delta]$ that contains $N_{\rm ev}$ eigenvalues. We do that by solving Eq.~\eqref{eq:Nev:exact} for $\delta$ using bisection.
    \item \textbf{Polynomial order.} Determine the polynomial order $K$ such that the interval $[\tilde{\lambda}-\delta, \tilde{\lambda}+\delta]$ (containing $N_\mathrm{ev}$ eigenvalues) lies above the cutoff $\Omega$ in the transformed spectrum. This is done by solving the exact expression in Eq.~\eqref{eq:condition:for:K} for $K$ using bisection.
    \item \textbf{Block Lanczos iteration with on-the-fly filtering.} Iterate the block Lanczos algorithm (block size $s$, at most $m$ block steps). At each step, the polynomial filter $P_{\tilde{\lambda}}^K(\tilde{H})$ is applied to the current Krylov block using the Clenshaw recurrence~\eqref{eq:clenshaw_chebyshev_prod}, requiring only $K$ sparse matrix-block multiplications (or mappings) with $\tilde{H}$. Full reorthogonalization is performed at every step.
    \item \textbf{Convergence check.} Periodically diagonalize the projected Lanczos matrix $H_L$ and evaluate the block residual norm. Stop when $\max_i \|B_{k+1} \tau_{ki}\| < \epsilon$.
    \item \textbf{Eigenpair extraction.} Transform the converged Lanczos eigenvectors back to the Hilbert-space basis, compute eigenvalues from Rayleigh quotients $E_i = \langle u_i | H | u_i \rangle$, and verify convergence via the residual norms $\|H u_i - E_i u_i\|$.
\end{enumerate}
The entire procedure requires storing only the Krylov blocks, the Lanczos matrix, and a small number of auxiliary vectors, with the sparse Hamiltonian $H$ accessed exclusively through matrix-vector products. This makes POLFED applicable to any problem for which an efficient sparse matrix-vector multiplication, or a specialized mapping encoding the action of $H$ on vectors, is available.

\section{Interface of \polfedjl}
\label{sec:interface}

In this section we describe the practical interface of \inh{Polfed.jl} and illustrate its usage through worked examples. The package is designed so that a minimal working calculation requires only a sparse Hamiltonian, a target energy, and the number of desired eigenpairs; all algorithmic parameters discussed in the previous section are set to sensible defaults but remain accessible for fine-tuning. We cover installation, basic usage, diagnostic reporting, target selection, the choice between standard and block Lanczos factorizations, automatic parameter optimization, custom Hamiltonian mappings, and GPU usage.

\subsection{Installation}
The package \inh{Polfed.jl} is installed through Julia's standard
package manager:
\begin{jlcodeblock}
using Pkg
Pkg.add("Polfed")
\end{jlcodeblock}
\noindent After installation, the package and the built-in quantum sun model~\eqref{eq:qsun}, used as a running example throughout this section, can be loaded with
\begin{jlcodeblock}
using Polfed
using Polfed.Models: qsun_hamiltonian
\end{jlcodeblock}

\subsection{Basic usage}

The main entry point of \inh{Polfed.jl} is the \inh{polfed()} function. In the simplest workflow, one first constructs a Hamiltonian matrix, then chooses the number of requested eigenpairs, the target energy, and an initial vector.
As a first example, we consider the quantum sun Hamiltonian:
\begin{jlcodeblock}
using LinearAlgebra
using Random

rng = MersenneTwister(1234)

L_loc = 7
L_grain = 3
g0 = 1.0
α = 0.55

use_U1_symmetry = false
Sz_sector = 0.0

mat = qsun_hamiltonian(
    L_loc, L_grain, g0, α;
    S=0.5, γ=1.0, w=0.5, hz=1.0, ζ=0.2,
    rng=rng, use_sparse=true, use_U1 = use_U1_symmetry, S_z = Sz_sector
)

x0 = rand(rng, size(mat, 1))
x0 ./= norm(x0)

howmany = 200
target = 0.0

vals, vecs = polfed(mat, x0, howmany, target)

\end{jlcodeblock}

\noindent Here, \inh{mat} is the Hamiltonian matrix, \inh{howmany} is the number of requested eigenpairs, \inh{target} specifies the target energy, and \inh{x0} is the initial normalized vector. Since the input \inh{x0} is a single vector (rather than a matrix), this call uses the standard Lanczos factorization; passing a $\mathcal{D} \times s$ matrix instead activates the block variant, as discussed in Sec.~\ref{sec:blocklanczosalgorithm}.
For further details on the model parameters considered, see Appendix~\ref{app:qsun}.

\subsection{Reports and diagnostics}

In addition to the computed eigenvalues and eigenvectors, \inh{Polfed.jl} can return a detailed report with diagnostic and timing information:
\begin{jlcodeblock}
vals, vecs, report = polfed(mat, x0, howmany, target)
display_report(report; convergence_details=true)
\end{jlcodeblock}

\noindent The report summarizes the most important aspects of the calculation. It contains information about the spectral transformation, such as the effective target, the exposed spectral interval, and the polynomial order, together with information about the factorization, including the number of converged eigenpairs, the convergence criteria, and the number of iterations performed. Timing data are also included, making the report useful for benchmarking different parameter choices or hardware configurations.

The report allows the user to check whether the chosen parameters lead to a reasonable polynomial order, whether the requested number of eigenpairs has converged, and which part of the computation dominates the runtime. This is especially useful for larger calculations, where the spectral transformation typically accounts for most of the runtime.

In the following subsections we use the report to illustrate the effect of block Lanczos and automatic optimization options.
The printed output can be adjusted to show more convergence details or to turn off the colors (useful to turn it off when one saves the report into the .txt file):
\begin{jlcodeblock}
display_report(
    report;
    color=true,
    convergence_details=true,
)
\end{jlcodeblock}

\subsection{Selection of the target energy}

The argument \inh{target} of \inh{polfed} controls where the polynomial filter is centered. This choice directly affects the polynomial order, convergence profile, and runtime of the calculation. Internally, \inh{Polfed.jl} works in rescaled spectral coordinates, see Eq.~\eqref{eq:rescaledMatrix}, so that the spectral interval is mapped to \([-1,1]\). The middle of the spectrum corresponds to $\tilde{E}=0$, and the point of maximal density of states is denoted by \(\tilde{E}_{\mathrm{maxdos}}\).

The interface supports the following target-selection modes:
\begin{itemize}
    \item \inh{target = E::Real} or \inh{target = (:unrescaled, E)} interprets the input as an unrescaled physical energy and converts it internally to the corresponding rescaled coordinate.
    \item \inh{target = (:rescaled, ϵ)} specifies the target directly in rescaled spectral coordinates.
    \item \inh{target = :middle} targets the middle of the spectrum and is equivalent to \inh{target =} \inh{(:rescaled, 0.0)}.
    \item \inh{target = :maxdos} targets \(\tilde{E}_{\mathrm{maxdos}}\), that is, the rescaled energy at which the density of states is maximal.
    \item \inh{target = (:offset, η)} selects a point displaced from \(\tilde{E}_{\mathrm{maxdos}}\) toward one of the spectral edges, with \(\eta=0\) corresponding to \(\tilde{E}_{\mathrm{maxdos}}\), \(\eta=1\) to \(+1\), and \(\eta=-1\) to \(-1\), with linear interpolation in between.
\end{itemize}

\begin{figure}[t]
    \centering
    \includegraphics[width=0.9\linewidth]{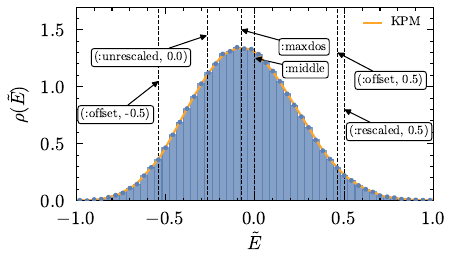}
    \caption{\textbf{Supported target specifications in \inh{Polfed.jl}.} Illustrated on the density of states of the rescaled spectral coordinates \(\tilde{E} \in [-1,1]\). The options \inh{:middle} and \inh{:maxdos} target the center of the spectrum and the point of maximal density of states, respectively. The form \inh{(:rescaled, ϵ)} specifies the target directly in rescaled coordinates, whereas \inh{(:unrescaled, E)} uses the physical energy before rescaling. \inh{(:offset, η)} selects a point displaced from $\tilde{E}_\mathrm{maxdos}$ toward one of the spectral edges. Data is shown for the $J_1$--$J_2$ model, see Eq.~\eqref{eq:j1j2}, for $L=20$ in the magnetization sector with $N_\uparrow=5$ up spins ($S^z_{\rm tot}=-5$).}
    \label{fig:dos_targets}
\end{figure}

\noindent Figure~\ref{fig:dos_targets} illustrates the different target specifications supported by \inh{Polfed.jl}. In particular, it shows the distinction between specifying a target directly in rescaled coordinates, specifying it in unrescaled physical energy units, and choosing it relative to the point of maximal density of states.

The following example demonstrates several of the supported target modes:
\begin{jlcodeblock}
x0 = rand(size(mat,1))
x0 ./= norm(x0)
howmany = 80

targets = (
    0.0,
    :maxdos,
    :middle,
    (:offset, 0.5),
    (:offset, -0.5),
    (:unrescaled, 0.0),
    (:rescaled, 0.5),
)

for t in targets
    vals, vecs = polfed(mat, x0, howmany, t)
end
\end{jlcodeblock}

\noindent Note that \inh{(:rescaled, 0.5)} and \inh{(:offset, 0.5)} are not equivalent. The former is measured from the center of the rescaled spectrum, whereas the latter is measured from \(\tilde{E}_{\mathrm{maxdos}}\) toward the right spectral edge.

The modes \inh{:maxdos} and \inh{(:offset, η)} rely on an estimated density of states. Consequently, the detected location of \(\tilde{E}_{\mathrm{maxdos}}\) can depend slightly on the DOS estimation settings. It is therefore advisable to inspect the reported target information when tuning these parameters.

\subsection{Lanczos and block Lanczos factorization}

The same interface supports both Lanczos and block Lanczos factorization. The distinction is determined automatically from the shape of the initial input \inh{x0}: a vector selects standard Lanczos, while a matrix activates block Lanczos with block size equal to the number of columns.

A minimal example is:
\begin{jlcodeblock}
howmany = 100
target = 0.0

# Vector input -> Lanczos factorization
x0_vec = rand(size(mat, 1))
x0_vec ./= norm(x0_vec)
vals_l, vecs_l = polfed(mat, x0_vec, howmany, target)

# Matrix input -> Block Lanczos factorization
x0_mat = rand(size(mat, 1), 4)
x0_mat = Matrix(qr(x0_mat).Q)
vals_b, vecs_b = polfed(mat, x0_mat, howmany, target)
\end{jlcodeblock}

\noindent In the first call, the input is a single normalized vector, so the algorithm uses the standard Lanczos factorization. In the second call, the input is a matrix with four columns, so block Lanczos is used with block size \(4\). The QR factorization is employed to construct a block of mutually orthonormal starting vectors, which is the recommended way to initialize a block run.

Block Lanczos applies the polynomial filter to several vectors simultaneously, which improves hardware utilization and is therefore especially beneficial for parallel execution on multi-core CPUs and GPUs (cf.\ Sec.~\ref{sec:gpu:polfed}).

However, a larger block size is not always better. Increasing block size can improve parallel efficiency by exposing more independent work during the mapping step, but it also changes the cost and memory footprint of the factorization itself. A useful rule of thumb is to choose the block size that satisfies:
\begin{equation}
\label{eq:howmany:block_size:ratio}
\frac{\texttt{howmany}}{\texttt{block\_size}} \gtrsim 100.
\end{equation}
If this ratio becomes too small, the block factorization typically requires many more Krylov iterations to converge, which can make the overall computation substantially slower. Equation~\eqref{eq:howmany:block_size:ratio} follows from the empirically obtained function below; it serves as an internal estimate for the required Krylov dimension:
\begin{jlcodeblock}
expectedkrylovdim(howmany::Int, blocksize::Int, η::Real) =
    ceil(Int64, (20.43*blocksize + 1.7*howmany)*η)
\end{jlcodeblock}
\noindent where \inh{η} is an overestimation factor. This estimate grows linearly with \inh{blocksize}, so choosing an unnecessarily large block can significantly increase the size of the projected problem and slow down convergence rather than improving it.

Increasing \inh{howmany} can, on the other hand, be beneficial. Requesting more eigenpairs typically decreases the required polynomial order, which reduces the cost of the spectral transformation and can lead to a net speedup.

However, \inh{howmany} cannot be increased indefinitely. A larger value also leads to a larger Krylov subspace, which increases both the memory needed to store the Krylov vectors (the dominant memory cost of the algorithm) and the size of the projected Lanczos matrix. Beyond some point, the memory requirements become prohibitive, and the cost of diagonalizing the projected matrix can additionally outweigh the benefit of the lower polynomial order. In practice, good performance is obtained by balancing these competing effects rather than maximizing either \inh{block\_size} or \inh{howmany} in isolation. This will be discussed later in Fig.~\ref{fig:tcpu:Nev} in Sec.~\ref{sec:benchmarks}.

When reporting is enabled, the factorization summary indicates which mode was used together with the block size and iteration count, making it straightforward to compare different configurations.

\subsection{Automatic optimization}

Beyond choosing the target and factorization type, one may ask whether the mapping itself (i.e., the action of the considered matrix on vectors), can be accelerated without writing a custom model-specific routine. In \inh{Polfed.jl}, this is achieved through the keyword \inh{optimize_mapping=true} inside \inh{MappingConfig}. Since POLFED is often memory-bound, reducing memory traffic in the mapping can lead to a substantial speedup.

A typical usage pattern is:
\begin{jlcodeblock}
mc = MappingConfig(optimize_mapping=true)

vals, vecs, report = polfed(mat, x0, howmany, target; mc=mc)
display_report(report)
\end{jlcodeblock}

\noindent The optimization of \inh{Polfed.jl} inspects the matrix structure and builds specialized routines that separate diagonal and off-diagonal contributions, compress repeated off-diagonal values, reduce memory traffic, and reuse the resulting routines in the spectral transformation and the Clenshaw recurrence. Since the mapping is applied many times during polynomial filtering, even moderate savings per call accumulate into a significant reduction of the total runtime.

This optimization is most effective for Hamiltonians with simple structure, e.g., when all off-diagonal elements share a common value or when only a few distinct off-diagonal values appear. In such cases, the generic sparse-matrix multiplication (explicitly accessing off-diagonal matrix elements) can be bypassed, significantly reducing memory access.

Automatic optimization thus serves as a convenient intermediate step between the generic matrix-based workflow and a handwritten custom mapping, see Sec.~\ref{sec:custom:mapping}. For structured Hamiltonians, it can often achieve the same speedup as a custom mapping while keeping the interface simpler.

A good practical workflow is to first use the ordinary matrix interface, then enable the \inh{optimize_mapping=true}, and only move to a custom mapping if further speedup is possible. The report makes it straightforward to compare optimized and non-optimized runs and to identify the remaining bottlenecks.

\subsection{Custom mappings and advanced usage}
\label{sec:custom:mapping}

So far, we have focused on the matrix-based interface, where the Hamiltonian is passed directly to \inh{Polfed.jl}. However, the POLFED algorithm, just like Lanczos itself, depends only on the action of the operator on a vector or a block of vectors. For this reason, \inh{Polfed.jl} also supports a function-based interface, where the user provides a custom mapping instead of an explicit matrix. This is the natural entry point for model-specific optimizations and for user-controlled parallelization.

In the first case, the package receives a matrix and applies it internally. In the second case, the user provides a mapping function \inh{f!} that performs the action of the operator on an input vector or block. The convention is that \inh{f!} operates in-place: \inh{f!(Y, X)} overwrites \inh{Y} with the result of applying the Hamiltonian to \inh{X}.

A minimal example is:
\begin{jlcodeblock}
f!(Y, X) = mul!(Y, mat, X)

vals, vecs, report = polfed(f!, x0, howmany, target)
\end{jlcodeblock}

\noindent The benefit of this interface is that one can write a custom mapping that avoids unnecessary memory access, exploits model-specific structure, or performs its own parallelization. This is especially useful for structured Hamiltonians, where the diagonal and off-diagonal parts can often be treated separately, or where the off-diagonal action can be generated directly from basis connectivity rules.

When a custom mapping is supplied, it is advisable to also provide its rescaled version. Otherwise, the rescaling must be applied at every step of the spectral transformation, introducing additional memory traffic. The spectral bounds must be supplied together with this custom rescaled mapping. If \inh{Emin} and \inh{Emax} are left unset, they will be re-estimated by an independent probe; consequently, the applied operator would no longer match the rescaling used to construct the filter, the density of states, and the target resolution. A rescaled wrapper can be constructed as:
\begin{jlcodeblock}
f!(Y, X) = mul!(Y, mat, X)

Emin, Emax = lanczos_extrema(mat)
ΔE = (Emax - Emin) / 2
Ec = (Emax + Emin) / 2

f!_rescaled = (Y, X) -> begin
    f!(Y, X)
    @. Y = Y / ΔE - (Ec / ΔE) * X
end

mc = MappingConfig(f!_rescaled=f!_rescaled, Emin=Emin, Emax=Emax)
\end{jlcodeblock}
\noindent Here, the \inh{lanczos_extrema()} function uses the built-in Lanczos algorithm to compute the minimum and maximum energies of the matrix. While the wrapper shown above has a pedagogical purpose, we note that its performance is not optimal: it costs two extra passes over the data on every Clenshaw step. For best performance, the rescaling should be fused directly into the custom mapping, or a rescaled matrix should be built once and passed directly.

Using a custom mapping also gives the user full control over parallelization. When the mapping already contains the desired threading or device-level execution, it is appropriate to combine the function-based interface with \inh{BlockWise()}, which hands the operator the whole Krylov block in a single call per Clenshaw step and leaves the decomposition entirely to the user's mapping. The alternative, \inh{VectorWise()}, instead opens one Clenshaw recurrence per Krylov vector and nests its own parallel region around a mapping that already threads. A typical setup is:
\begin{jlcodeblock}
mc = MappingConfig(
    parallel_strategy=BlockWise(),
    f!_rescaled=f!_rescaled,
    Emin=Emin,
    Emax=Emax,
)

vals, vecs, report = polfed(f!, x0, howmany, target; mc=mc)
\end{jlcodeblock}

\noindent In many structured models, a custom mapping can provide the best performance, since it allows the user to tailor the mapping directly to the underlying Hamiltonian and to avoid unnecessary overhead in the repeated spectral transformation.

The function-based interface also serves as the starting point for GPU execution (see below), while retaining the same \inh{Polfed.jl} call signature.

\subsection{GPU usage}
Using \inh{Polfed.jl} on GPUs is straightforward. The user does not need to write GPU-specific code. One only has to move the Hamiltonian and initial vectors to the device (GPU), after which the computations are carried out there automatically.

A typical workflow begins by constructing the Hamiltonian and the initial vector on the CPU, and then transferring them to the device:
\begin{jlcodeblock}
using CUDA

mat_gpu = CUDA.CUSPARSE.CuSparseMatrixCSR(mat)

x0 = rand(size(mat, 1))
x0 ./= norm(x0)
x0_gpu = CuArray(x0)
\end{jlcodeblock}

\noindent GPU support is provided by a package extension that stays dormant until CUDA is
loaded, so nothing extra has to be installed at the package level---the \inh{using CUDA}
above is what activates the device code paths.

Once the arrays are on the GPU, the \inh{polfed} call is unchanged:
\begin{jlcodeblock}
# On a GPU the strategy always resolves to BlockWise: the device owns the
# decomposition. Setting it explicitly only documents that intent.
mc = MappingConfig(parallel_strategy=BlockWise())

vals_gpu, vecs_gpu, report_gpu = polfed(
    mat_gpu,
    x0_gpu,
    howmany,
    target;
    mc=mc,
)
\end{jlcodeblock}

\noindent After the data have been placed on the GPU, the mapping and the associated linear algebra operations are performed there. GPU execution naturally pairs with \inh{BlockWise()}, which hands the operator the whole Krylov block in one call per Clenshaw step and leaves the decomposition to the device; splitting matrix-matrix multiplication into multiple matrix-vector multiplications would only add overhead. This strategy is the default for both CPU and GPU arrays.

Further optimizations are possible. For example, considering the quantum random energy model (QREM), see Eq.~\eqref{eq:qrem}, one can construct a very efficient mapping that minimizes the data transfer between global GPU memory and the processing units:
\begin{jlcodeblock}
function qrem_map_kernel(
    Y::CuDeviceVector{T}, X::CuDeviceVector{T}, 
    L::Int, diags::CuDeviceVector, Γ::Real, basis_length::Int
) where {T<:Real}
    i = (blockIdx().x - 1) * blockDim().x + threadIdx().x

    # Ensures the thread index does not exceed 
    # the vector size to prevent out-of-bounds access
    @inbounds if i <= basis_length 
        offdiag_val = 0.0 

        for l in 0:L-1
            newstate = (i-1) ⊻ (1 << l) 
            row = newstate + 1 # because of 1-based enumeration of vectors
            offdiag_val += X[row] 
        end     
        
        Y[i] = diags[i]*X[i] + offdiag_val*Γ #store the result to final array
    end

    return nothing
end
\end{jlcodeblock}

\noindent This mapping was used to benchmark the QREM code, and to compare it versus the CPU architecture in Fig.~\ref{fig:gpu:bench} in Sec.~\ref{sec:bench:gpu}

GPU calculations can use both the standard Lanczos and the block Lanczos factorization, depending on the shape of \inh{x0\_gpu}. In practice, block input is often preferable on GPUs, since applying the mapping to several vectors at once improves device utilization (cf.\ Sec.~\ref{sec:gpu:polfed}).

A block turns the transform's matrix-vector products into matrix-matrix products, which
raises the arithmetic intensity and gives the device far more concurrent work. The block is
constructed on the CPU and moved to the GPU like any other array:

\begin{jlcodeblock}
x0_gpu_block = CuArray(Matrix(qr(rand(size(mat, 1), 8)).Q))

vals_gpu, vecs_gpu, report_gpu = polfed(mat_gpu, x0_gpu_block, howmany, target)
\end{jlcodeblock}

The same considerations apply to custom mappings. If the user provides a GPU-resident mapping, it should follow the same in-place convention \inh{f!(Y, X)}. As on the CPU, it is advisable to provide the rescaled version of the mapping, in order to avoid additional memory traffic from repeated on-the-fly rescaling, together with the spectral bounds \inh{Emin} and \inh{Emax} it was built from.

In summary, GPU support in \inh{Polfed.jl} is simple: construct the data in the usual way, move the arrays to the GPU, and call \inh{polfed}; the computation proceeds on the device.

\section{Benchmarks}
\label{sec:benchmarks}
In this section we benchmark the performance of POLFED against shift-and-invert, and compare our CPU and GPU implementations. We first identify the regimes in which POLFED is favorable compared to shift-and-invert (Sec.~\ref{sec:polfed:vs:si}), then determine which parts of POLFED dominate the runtime (Sec.~\ref{sec:bench:parts}) and how the cost scales with the number of requested eigenpairs (Sec.~\ref{sec:bench:nev}), and finally compare CPU and GPU hardware (Sec.~\ref{sec:bench:gpu}).

\paragraph{Computational details.}
\label{para:comp:details}
All calculations use double-precision (64-bit) floating-point arithmetic, since single precision does not provide sufficient accuracy, as also argued in Ref.~\cite{Pietracaprina2018}. POLFED terminates once both the Ritz residual criterion~\eqref{eq:blocklan_criterion} and the eigenpair criterion $\lVert Hu_i - E_i u_i \rVert < \varepsilon_{\text{eig}}$ are satisfied; for both we use the default tolerances, $10^{-14}$ and $10^{-9}$, respectively.

The POLFED benchmarks were obtained with \inh{Polfed.jl} under Julia~1.11~\cite{Bezanson17julia}. The CPU runs used a dual-socket AMD EPYC 9354 node ($2\times32$ cores, Zen~4, 3.25~GHz base clock), except for those in Fig.~\ref{fig:gpu:bench}, which used a dual-socket AMD EPYC 7H12 node ($2\times64$ cores, Zen~2, 2.6~GHz base clock). The GPU runs were performed with CUDA~13 on an NVIDIA GH200 Grace Hopper Superchip with 144~GB of HBM3e memory.

The shift-and-invert benchmarks were obtained with a C++ implementation based on PETSc~\cite{petsc-manual, petsc-efficient} and SLEPc~\cite{Hernandez05slepc}, parallelized across nodes with MPI and using MUMPS~\cite{Amestoy01mumps, Amestoy06mumps} for the sparse LU factorization. These runs used dual-socket Intel Xeon Platinum 8268 nodes ($2\times24$ cores at 2.90~GHz, four NUMA domains of 12 cores each, SMT disabled).

\subsection{POLFED vs Shift-And-Invert}
\label{sec:polfed:vs:si}
\begin{figure}[b!]
    \centering
    \includegraphics[width=0.99\columnwidth]{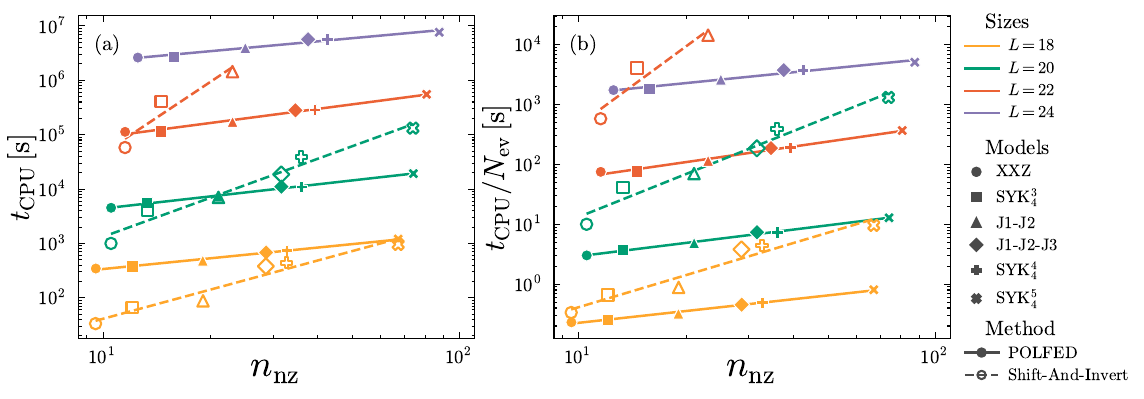}

    \caption{\textbf{POLFED vs shift-and-invert, CPU time.} CPU-time comparison between POLFED and shift-and-invert benchmarks. Panel (a) shows the total CPU time as a function of $n_{\rm nz}$, the number of nonzero off-diagonal matrix elements per row. Panel (b) shows the corresponding CPU time per targeted eigenvalue. Colors indicate the system size $L$, while marker shapes distinguish the different model classes (the XXZ model from Eq.~\eqref{eq:xxz}, the $J_1$--$J_2$ model from Eq.~\eqref{eq:j1j2}, the $J_1$--$J_2$--$J_3$ model from Eq.~\eqref{eq:j1j2j3}, and the $\text{SYK}_4^d$ model from Eq.~\eqref{eq:syk4_d}). In these benchmarks, POLFED targeted $1500$ eigenvalues, whereas shift-and-invert targeted only $100$ eigenvalues due to memory constraints. All simulations used $16$ CPUs with Block Lanczos factorization (block size was $16$). The shift-and-invert dataset is incomplete because, for the largest systems, the available cluster memory was insufficient to store the LU factors. For POLFED, matrix multiplication was performed using Julia's standard \inh{LinearAlgebra} routine \inh{mul!()}. Using the optimized mapping for models with constant off-diagonal values would provide an additional speedup of approximately a factor of $3$.}
    \label{fig:polfed:vs:si}
\end{figure}

We begin by comparing \inh{Polfed.jl} with the shift-and-invert algorithm widely used for computing mid-spectrum eigenpairs of many-body Hamiltonians~\cite{Luitz15, Serbyn17, Mace18, Pietracaprina2018, Suntajs20, Colbois23, Colbois24, Laflorencie25}. As discussed in Sec.~\ref{Methods:for:targeting:central:part:of:the:spectrum}, the key difference between POLFED and shift-and-invert is that the latter requires a sparse LU factorization of the shifted Hamiltonian, whereas POLFED applies the polynomial filter on the fly using only sparse matrix-vector products.

The comparison is shown in Fig.~\ref{fig:polfed:vs:si}, where we plot the CPU times of POLFED and shift-and-invert across different model families as a function of the number of off-diagonal matrix elements per row. Letting $N_{\rm off}$ denote the total number of off-diagonal matrix elements, this per-row average is given by $n_\mathrm{nz} = N_{\rm off}/\mathcal{D}$. In these benchmarks, POLFED targeted \(1500\) eigenvalues, whereas shift-and-invert targeted only \(100\) eigenvalues due to the substantially larger memory requirements of the factorization-based approach, which is why we show total CPU time in Fig.~\ref{fig:polfed:vs:si}(a) and CPU time per eigenvalue in Fig.~\ref{fig:polfed:vs:si}(b). Moreover, the shift-and-invert dataset is not complete for all cases, since for the largest problems the available cluster memory was insufficient to store the factorized matrices. This already illustrates an important practical distinction between the two methods.

The cost of shift-and-invert is dominated by the factorization of the shifted Hamiltonian, which suffers from fill-in: the LU factors are much denser than the original sparse Hamiltonian, leading to large memory requirements. By contrast, POLFED never constructs or factorizes a transformed operator explicitly. Instead, its cost is dominated by repeated sparse matrix-vector or sparse matrix-matrix multiplications during the spectral transformation. As a result, POLFED remains applicable in regimes where shift-and-invert is limited by memory.

Compared to shift-and-invert, POLFED offers several practical advantages:
\begin{itemize}
    \item \textbf{Low memory consumption.} The transformed operator is never constructed or factorized explicitly, which greatly reduces the memory footprint.
    \item \textbf{No fill-in problem.} Since no sparse factorization is performed, POLFED avoids the fill-in that makes shift-and-invert increasingly expensive for less sparse Hamiltonians.
    \item \textbf{Favorable computational scaling.} The dominant operations are repeated sparse matrix-vector or sparse matrix-matrix multiplications, whose cost grows more mildly with connectivity than the factorization step in shift-and-invert.
    \item \textbf{No need for distributed factorization routines.} POLFED does not require multi-node MPI parallelization for sparse factorizations.
    \item \textbf{Simpler software stack.} Unlike shift-and-invert workflows that often depend on external packages such as PETSc and SLEPc, \inh{Polfed.jl}{} is self-contained.
    \item \textbf{Natural compatibility with GPUs.} Since the cost is concentrated in repeated sparse mappings, POLFED is well suited for GPU acceleration, as shown in Sec.~\ref{sec:bench:gpu}.
    \item \textbf{Model-specific optimizations.} The mapping step can be specialized for particular Hamiltonian structures, allowing optimizations beyond a generic sparse-matrix implementation.
\end{itemize}

\subsection{Runtime breakdown of POLFED}
\label{sec:bench:parts}

\begin{figure}[t]
    \centering
    \includegraphics[width=0.99\columnwidth]{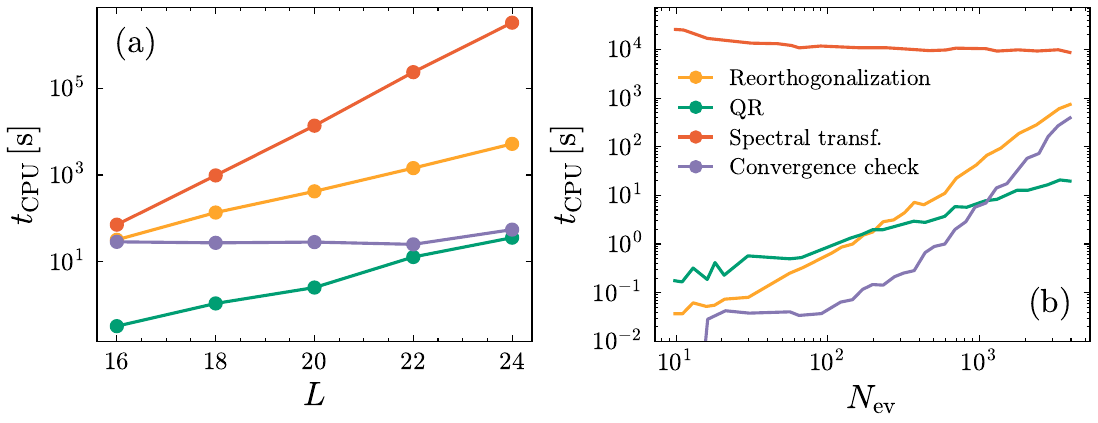}
    \caption{\textbf{Decomposition of the POLFED CPU time.} It is decomposed into its main computational parts: spectral transformation, QR decomposition, reorthogonalization, and convergence checking. Calculations were performed for the XXZ model, see Eq.~\eqref{eq:xxz}, in the zero-magnetization sector. Panel (a) shows the scaling with system size \(L\) for $N_{\rm ev}=1500$ targeted eigenvalues with block size $16$. Panel (b) shows the scaling with the number of targeted eigenvalues $N_{\rm ev}$ at fixed system size \(L=22\). Matrix multiplication was performed with Julia's standard \inh{LinearAlgebra} function \inh{mul!()}. Employing the optimized mapping for models with constant off-diagonal values would yield an additional speedup of approximately a factor of \(3\).}
    \label{fig:tcpu:parts}
\end{figure}

Here, we decompose the total runtime of a \inh{Polfed.jl} run into its main components. In Fig.~\ref{fig:tcpu:parts} we show the CPU time spent on the spectral transformation, QR decomposition, reorthogonalization, and convergence checking.

As expected, the spectral transformation dominates the total cost, since POLFED applies it repeatedly within the block Lanczos procedure. This means that improvements in the mapping step translate directly into a proportional reduction of the total runtime. For large numbers of requested eigenvalues, however, convergence checking becomes increasingly important because it requires repeatedly diagonalizing the projected Lanczos matrix.

To interpret these results quantitatively, we summarize the leading-order cost of each step of the block Lanczos iteration with block size $s$, $m$ block steps, polynomial order $K$, Hilbert-space dimension $\mathcal{D}$, and $n_\mathrm{nz}$ nonzero elements per row of $H$:
\begin{itemize}
    \item \textbf{Spectral transformation.} Each block step applies the polynomial filter via $K$ sparse matrix-block products, each costing $\mathcal{O}(n_\mathrm{nz}\,\mathcal{D}\,s)$. The total cost over $m$ steps is $\mathcal{O}(M\,K\,n_\mathrm{nz}\,\mathcal{D})$.
    \item \textbf{Reorthogonalization.} At step $k$, the new block is orthogonalized against all $k$ previous blocks, costing $\mathcal{O}(k\,\mathcal{D}\,s^2)$. Summing over $m$ steps gives $\mathcal{O}(M^2\,\mathcal{D})$.
    \item \textbf{QR decomposition.} Each step requires a QR factorization of a $\mathcal{D}\times s$ block at cost $\mathcal{O}(\mathcal{D}\,s^2)$, giving $\mathcal{O}(M\,\mathcal{D}\,s)$ in total.
    \item \textbf{Convergence check.} At each convergence check, one diagonalizes the projected block-Lanczos matrix of dimension $M=ms$. If treated as a dense symmetric matrix, this costs $\mathcal{O}(M^3)=\mathcal{O}(m^3s^3)$ per check. Since checks are performed periodically, the accumulated cost depends on the checking frequency and can reach $\mathcal{O}(mM^3)=\mathcal{O}(m^4s^3)$ in the worst case. In the scalar Lanczos case, $s=1$, the projected matrix is tridiagonal, and specialized symmetric tridiagonal eigensolvers can reduce the diagonalization cost to $\mathcal{O}(M^2)$~\cite{Parlett1998SymmetricEigenvalueProblem,DhillonParlett2004MRRR,Demmel2008TridiagonalEigensolvers}.
\end{itemize}
For a fixed number of targeted eigenvalues $N_\mathrm{ev}$ and block size $s$, the Krylov dimension $M$ is approximately constant, so the spectral transformation cost grows as $\mathcal{O}(K\,n_\mathrm{nz}\,\mathcal{D})$. Since $K$ itself grows exponentially with $L$ (at fixed $N_\mathrm{ev}$, cf.\ Sec.~\ref{sec:polynomial:spectral:filtering}), the spectral transformation dominates the scaling with system size. Conversely, increasing $N_\mathrm{ev}$ at fixed $L$ reduces $K$ but enlarges the Krylov subspace, shifting the balance toward reorthogonalization and convergence checking.

\subsection{Scaling with the number of requested eigenpairs}
\label{sec:bench:nev}

\begin{figure}[htb]
    \centering
    \includegraphics[width=\columnwidth]{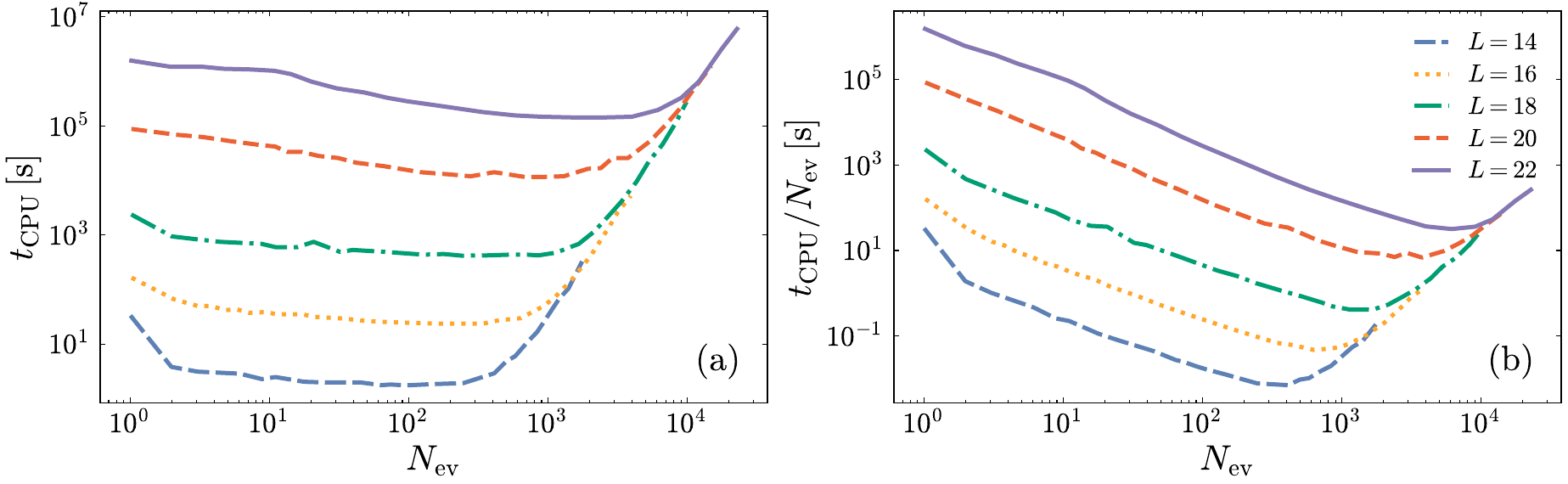}
    \caption{\textbf{CPU time of POLFED for different numbers of eigenpairs.} CPU time of POLFED as a function of the number of requested eigenpairs \(N_{\rm ev}\) for the XXZ model, Eq.~\eqref{eq:xxz}, in the zero-magnetization sector. Colors indicate the system size \(L\). Panel (a) shows the total CPU time \(t_{\rm CPU}\), while panel (b) shows the CPU time per requested eigenpair, \(t_{\rm CPU}/N_{\rm ev}\). Matrix multiplication was performed with Julia's standard \inh{LinearAlgebra} function \inh{mul!()}, and for all benchmarks block Lanczos factorization was used with block size $8$. Employing the optimized mapping for models with constant off-diagonal values would yield an additional speedup of approximately a factor of $3$. }
    \label{fig:tcpu:Nev}
\end{figure}

As discussed in the scaling analysis of the previous subsection, increasing $N_\mathrm{ev}$ has two competing effects: the polynomial order $K$ decreases, reducing the cost of the spectral transformation, while the Krylov subspace grows, increasing the cost of reorthogonalization, QR decomposition, and convergence checking. Figure~\ref{fig:tcpu:Nev}(a) illustrates the resulting trade-off: for small $N_\mathrm{ev}$, the spectral transformation dominates, and the total CPU time decreases as $N_\mathrm{ev}$ increases. For sufficiently large $N_\mathrm{ev}$, the dense linear-algebra steps associated with the growing Krylov subspace become more important. Figure~\ref{fig:tcpu:Nev}(b) shows how much CPU time is needed per eigenvalue. Further information about the interplay of spectral transformation and final diagonalization (convergence checking) is given in Appendix \ref{app:optimal:number:of:nev}.

\subsection{CPU vs GPU implementation}
\label{sec:bench:gpu}

\begin{figure}[b!]
    \centering
    \includegraphics[width=0.9\columnwidth]{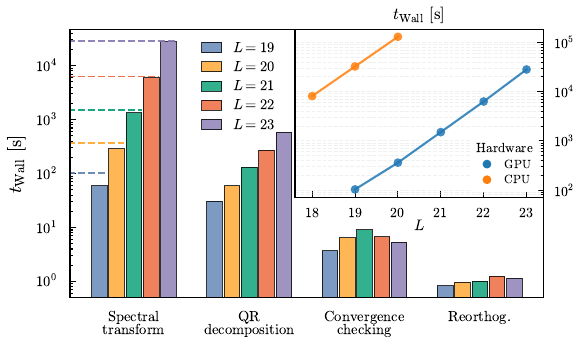}
    \caption{\textbf{Benchmarks of the GPU implementation of POLFED}. Main panel: breakdown of the total wall time into the main computational components for different system sizes $L$ on the GPU node. Inset: comparison of the total wall time of the GPU and CPU implementations. The benchmarks were performed for the QREM model, Eq.~\eqref{eq:qrem}. In all runs, $1500$ eigenpairs were targeted at the maximum of the density of states (target \inh{:maxdos}, see Fig.~\ref{fig:dos_targets}), except for the largest GPU run, where $750$ eigenpairs were targeted (because of the memory constraints). For calculations on CPU architecture, 16 cores were used with Block Lanczos factorization (with block size $16$), while on GPU we used Lanczos factorization (block size 1). The mapping step was implemented by exploiting the structure of the off-diagonal matrix elements; see the discussion in Sec.~\ref{sec:custom:mapping} about the optimized mapping. Dashed colored lines on the right side of the main panel represent the total wall time of the program, indicating that the spectral transformation remains the most time-demanding part of the program. More details about the CPU and GPU architecture can be found in Sec.~\ref{para:comp:details}.}
    \label{fig:gpu:bench}
\end{figure}

Finally, we compare the CPU and GPU implementations of POLFED. All GPU benchmarks were done on a single GPU. The results are summarized in Fig.~\ref{fig:gpu:bench}, which shows both the runtime breakdown of the GPU implementation and the total wall-time comparison between the CPU and GPU calculations. Comparison was done for the quantum random energy model (QREM), see Eq.~\eqref{eq:qrem}. It is important to note that this model does not possess any symmetries and therefore spans the full Hilbert space of dimension $\mathcal{D}=2^L$.

Since the spectral transformation dominates the runtime and consists of memory-bandwidth-limited sparse operations (cf.\ Sec.~\ref{sec:gpu:polfed}), one expects a substantial speedup from GPU offloading. In these benchmarks, \(1500\) eigenpairs were targeted at the maximum of the density of states. The main panel of Fig.~\ref{fig:gpu:bench} shows the decomposition of the total wall time into the main computational components for different system sizes \(L\), while the inset compares the total wall time of the GPU and CPU implementations. Once the spectral transformation is accelerated on the GPU, the relative weight of the remaining steps (reorthogonalization, QR, convergence checking) shifts, providing a clear picture of where further optimization effort should be directed.

The inset of Fig.~\ref{fig:gpu:bench} demonstrates the most striking result of this comparison: for \(L=20\), the GPU implementation reduces the total wall time from roughly \(2000\)~minutes on \(16\) CPU cores to approximately \(6\)~minutes on a single GPU, corresponding to a speedup of more than two orders of magnitude. We stress that this comparison is between two different hardware architectures (16 CPU cores versus a single GPU), and therefore does not represent a fixed-hardware scaling test. Nevertheless, it highlights the practical impact of GPU offloading for POLFED: calculations that would require days of single-core CPU time become feasible in minutes on a modern GPU. This is a direct consequence of the algorithmic structure of POLFED, where the dominant cost reduces to repeated sparse matrix-block multiplications, an operation for which GPUs provide vastly higher memory bandwidth and arithmetic throughput than the CPU architecture.

\section{Conclusion} \label{sec:conclusion}
 
We have presented \inh{Polfed.jl}, an open-source Julia package implementing the Polynomially Filtered Exact Diagonalization (POLFED) algorithm for computing mid-spectrum eigenvalues and eigenvectors of large sparse matrices, with a focus on quantum many-body Hamiltonians. The paper provides a self-contained description of the method, covering the underlying Krylov-subspace theory (Sec.~\ref{sec:krylov:space:methods}), spectral transformations and their trade-offs (Sec.~\ref{Methods:for:targeting:central:part:of:the:spectrum}), the complete POLFED algorithm including on-the-fly polynomial filtering via the Clenshaw recurrence (Sec.~\ref{sec:POLFED}), the user-facing interface of \inh{Polfed.jl} (Sec.~\ref{sec:interface}), and systematic benchmarks against shift-and-invert as well as CPU--GPU comparisons (Sec.~\ref{sec:benchmarks}).
 
The key advantage of POLFED over the widely used shift-and-invert approach is that it avoids sparse matrix factorization entirely. Instead, the polynomial spectral filter is evaluated on the fly through repeated sparse matrix-vector products, preserving the sparsity of the Hamiltonian and drastically reducing memory requirements. This makes it possible to reach system sizes that are inaccessible to shift-and-invert due to the fill-in associated with LU decomposition. The algorithmic structure of POLFED, dominated by sparse matrix-block multiplications, is naturally suited for GPU acceleration: our benchmarks demonstrate speedups of two orders of magnitude when moving from a 16 CPU cores to a single high-end GPU.

The \inh{Polfed.jl} package is designed to be both accessible and flexible. In its simplest usage, a calculation requires only a sparse Hamiltonian, a target energy, and the number of desired eigenpairs, with all algorithmic parameters set to sensible defaults. At the same time, the package exposes a function-based interface for custom Hamiltonian mappings, automatic optimization of the spectral transformation for structured matrices, and full GPU support, allowing advanced users to tailor the computation to their specific problem.

POLFED has already been employed to obtain numerically exact mid-spectrum eigenpairs across a broad range of quantum systems, establishing it as a well-tested tool in the field. Applications include the study of thermalization and ergodicity-breaking phenomena in paradigmatic quantum many-body Hamiltonians such as the disordered XXZ and $J_1$--$J_2$ spin chains~\cite{Sierant20polfed, Aramthottil25njp, Falcao26xxz}, quasiperiodic XXZ models~\cite{Aramthottil21quasiperiodic, Falcao24qp}, bond-disordered XXZ chains~\cite{Aramthottil24bond, Prodius26}, the clean and disordered mixed-field Ising model~\cite{Turkeshi25ising}, and the disordered PXP model~\cite{Sierant21pxp}. Beyond conventional spin chains, POLFED has been applied to the quantum sun model and its variants~\cite{Pawlik24mobilityEdge, Pawlik25unconventional,Swietek25scaling}, to Anderson localization on random graphs~\cite{Sierant23randomgraphs, Vanoni24randomgraphs}, and to higher-dimensional Anderson models~\cite{Vanoni25highdim,Jiricek26Universal}. This wide range of successful applications demonstrates the robustness and versatility of the method. With the release of \inh{Polfed.jl}, we aim to make this capability readily accessible to the broader community and to facilitate future large-scale studies of thermalization, ergodicity breaking, and other non-equilibrium phenomena in quantum many-body systems, and beyond, for any problem in which mid-spectrum eigenpairs of large sparse matrices are of interest.

An important direction for future development concerns periodically driven (Floquet) systems. The idea of using polynomial spectral transformations for the partial diagonalization of large Floquet unitary operators was put forward in Refs.~\cite{Luitz21floquet, Sierant23floquet}, and has already been employed to study avalanches and many-body resonances in Floquet and Hamiltonian strongly disordered systems~\cite{Morningstar22avalanches}, as well as anomalous transport in U(1)-symmetric quantum circuits~\cite{Summer26transport}. The existing implementations, however, are tailored to specific models such as the kicked Ising chain~\cite{Sierant23floquet}, where the Floquet operator factorizes into a product of simple unitaries that can be applied efficiently. Developing a general-purpose method capable of handling generic Floquet operators, without relying on a particular factorization of the time-evolution operator, is an interesting open challenge and a natural next step for the development of polynomial spectral transformations and their application to non-equilibrium many-body dynamics. We leave this question for future work.

\section*{Acknowledgements}
R.P. thanks Domen Vaupotič for his help during the development of the \inh{Polfed.jl} package. R.P. also thanks Luka Leskovec for discussions regarding GPU implementation and optimization procedures, and Simon Jiricek and Kohei Ogane for useful discussions and comments.
We thank Fabien Alet for technical assistance with configuring the shift-and-invert code on the HPC cluster. R.P., R.S., M.H., J.Š. and L.V. acknowledge support from the Slovenian Research and Innovation Agency (ARIS), Research core funding Grants No.~P1-0044, N1-0273, J1-50005 and N1-0369, as well as the Consolidator Grant Boundary-101126364 of the European Research Council (ERC). M. H. acknowledges partial support from the Polish National Agency for Academic Exchange (NAWA)’s Ulam Programme (project BNI/ULM/2024/1/00124).
The work of K.P and J.Z. was funded by the National Science Centre, Poland, project  2021/43/I/ST3/01142 -- OPUS call within the WEAVE programme.
They acknowledge Polish high-performance computing infrastructure PLGrid (HPC Center: ACK Cyfronet AGH) for providing computer facilities and support within computational grant no. PLG/2025/018400.
P.S. acknowledges fellowship within the “Generación D” initiative, Red.es, Ministerio para la Transformación Digital y de la Función Pública, for talent attraction (C005/24-ED CV1), funded by the European Union NextGenerationEU funds through PRTR.
We gratefully acknowledge the High Performance Computing Research Infrastructure Eastern Region (HCP RIVR) consortium~\cite{vega1} and European High Performance Computing Joint Undertaking (EuroHPC JU)~\cite{vega2} for funding this research by providing computing resources of the HPC system Vega at the Institute of Information sciences~\cite{vega3}.

\appendix

\section{Filtered-subspace diagonalization algorithms}\label{app:feast:dacp}

\subsection{FEAST method} \label{sec:feast:method}

Over the years, several alternative eigensolvers have been proposed.
One such example is the FEAST eigensolver, introduced by Polizzi~\cite{Polizzi2009FEAST,Tang2014FEAST}, which has so far been used in only a few studies of quantum many-body physics~\cite{Kumar2022FEAST,Vogl2019Flow,Schaefer2020Spectral}. FEAST differs substantially from the Krylov-based methods discussed above. Rather than constructing a Krylov subspace, it employs a rational spectral filter that approximately projects onto the eigenspace associated with a prescribed energy interval. The corresponding spectral projector can be written as
\begin{equation}
    \mathcal{P}=\frac{1}{2\pi i}\oint_{\mathcal C}(z\mathds{1}-H)^{-1}\,\mathrm{d}z,
\end{equation}
where the contour $\mathcal C$ encloses the desired part of the spectrum. The contour integral is evaluated numerically using a finite set of quadrature points $z_j\in\mathcal C$,
\begin{equation}
    V_{\mathrm{FEAST}}
    =\mathcal{P}Q
    \approx
    \sum_j w_j (z_j\mathds{1}-H)^{-1}V,
\end{equation}
where $V$ is an initial block of trial vectors and $w_j$ are the corresponding quadrature weights. Rather than inverting each shifted matrix, one solves, for every quadrature point, the shifted linear system,
\begin{equation}
    (z_j\mathds{1}-H)X_j=V,
\end{equation}
so that the filtered subspace is obtained from the weighted sum,
\begin{equation}
    V_{\mathrm{FEAST}}
    \approx
    \sum_j w_j X_j.
\end{equation}
After orthogonalization, the original Hamiltonian is projected onto this much smaller subspace,
\begin{equation}
    H_{\mathrm{sub}}
    =
    V_{\mathrm{FEAST}}^{\dagger}
    H
    V_{\mathrm{FEAST}},
\end{equation}
and the resulting reduced eigenvalue problem is diagonalized.

Despite its elegant formulation and excellent parallelizability, FEAST can become computationally demanding for large many-body Hamiltonians. Its main bottleneck is the solution of several shifted linear systems which, when performed with a sparse direct factorization, leads to very large memory requirements because of the fill-in phenomena. Iterative variants reduce the memory cost, but their efficiency depends strongly on the convergence of the inner linear solver and on the availability of suitable preconditioners. In this sense, FEAST and shift-and-invert are closely related, since both rely on the resolvent $(z\mathds{1}-H)^{-1}$. An advantage of the block and interval-based formulation is that all eigenstates within the selected energy window are obtained simultaneously, including degenerate and nearly degenerate states, provided that the chosen subspace dimension is sufficiently large.

\subsection{Dual applications of Chebyshev polynomials (DACP) method}\label{sec:dacp}

Another approach specifically designed for computing a large number of eigenstates near the center of the spectrum is the DACP method, introduced in Ref.~\cite{Guan2021Dual}. In contrast to FEAST, DACP does not require the solution of shifted linear systems, but relies entirely on repeated matrix-vector multiplications, and can therefore be implemented in a matrix-free fashion.

The method uses Chebyshev polynomials at two distinct stages. First, a high-order Chebyshev filter is applied to an initial random vector $|\theta_1\rangle$ in order to suppress contributions from eigenstates outside the desired central energy window,
\begin{equation}
    |\theta_{\mathrm{f}}\rangle
    =
    F(H)|\theta_1\rangle
    \approx
    \sum_{E_n\in\mathcal I}
    c_n |u_n\rangle,
\end{equation}
where $\mathcal I$ denotes the target spectral interval and $F(H)$ is a polynomial filter constructed from Chebyshev polynomials. The resulting state therefore consists predominantly of eigenstates belonging to the desired energy window.
Since a single filtered vector is not sufficient to resolve all eigenstates within the interval, Chebyshev polynomials are applied a second time to generate a set of linearly independent vectors,
\begin{equation}
    |\Theta\rangle
    =
    T_k(\tilde{H})|\theta_{\mathrm{f}}\rangle,
\end{equation}
where $\tilde{H}$ denotes the rescaled Hamiltonian, see Eq.~\eqref{eq:rescaledMatrix}, and $T_k$ is the Chebyshev polynomial of order $k$. These vectors approximately span the eigenspace associated with the target energy interval. The Hamiltonian and overlap matrices are then constructed within this generally nonorthogonal subspace,
\begin{equation}
    \left[H_{\mathrm{sub}}\right]_{ij} = \langle \Theta_i|H|\Theta_j\rangle,
    \qquad
    S_{ij} = \langle \Theta_i|\Theta_j\rangle,
\end{equation}
and the desired eigenvalues are obtained from the reduced generalized eigenvalue problem,
\begin{equation}
    H_{\mathrm{sub}} B = S B \Lambda\,.
\end{equation}
The main advantage of DACP is that it avoids sparse matrix factorizations altogether and requires only the action of $H$ on a vector, which results in comparatively low memory requirements for large many-body Hamiltonians. Its main limitation is that the resulting eigenvalues are not converged to numerical precision, which is necessary for some applications, like computing the level spacings of the model. Apart from that, the method is regarded as fast and memory efficient in comparison with the shift-and-invert method of Sec.~\ref{sec:shift:and:invert}: in the benchmarks of Ref.~\cite{Guan2021Dual}, it was reported to be up to $30$ times faster while requiring substantially less memory.

\section{Accuracy and reliability of the method}
\label{app:accuracy:and:reliability}

\begin{figure}[b!]
    \centering
    \includegraphics[width=0.999\columnwidth]{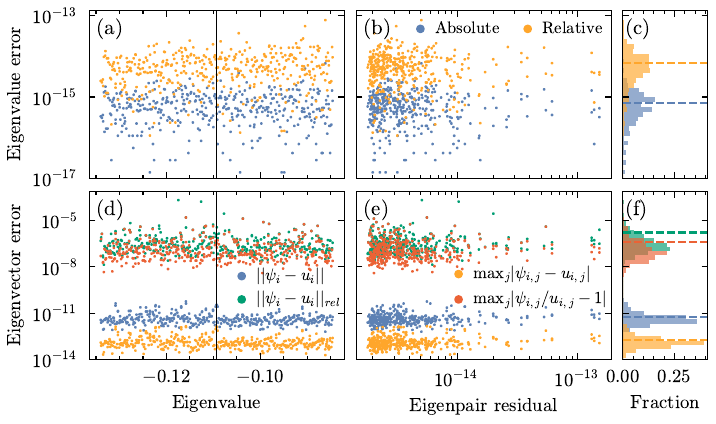}
    \caption{\textbf{Errors of the eigenvalues and eigenvectors obtained with POLFED.} The first row (panels (a)--(c)) shows the absolute and relative eigenvalue errors, $|\varepsilon_i - E_i|$ and $|\varepsilon_i/E_i - 1|$. The second row (panels (d)--(f)) shows two measures of the eigenvector error, the distance to the exact eigenstate $\|\psi_i - u_i\|$ and the maximal coefficient error $\max_j |\psi_{j,i} - u_{j,i}|$, each in its absolute and relative form. In the first column (panels (a) and (d)) the errors are plotted as a function of the eigenvalue, with the solid vertical line marking the target energy of the run. The second column (panels (b) and (e)) shows them as a function of the eigenpair residual $\|H\psi_i - \varepsilon_i\psi_i\|$, and the third column (panels (c) and (f)) shows their distributions, with dashed lines indicating the mean values. Results are shown for the disordered XXZ model in the zero-magnetization sector with $L=18$, obtained with $N_{\rm ev}=1000$ requested eigenpairs and block size $4$. Errors are calculated with respect to the data obtained from full exact diagonalization.}
    \label{fig:error:eigenpair}
\end{figure}

\begin{figure}[b!]
    \centering
    \includegraphics[width=0.999\columnwidth]{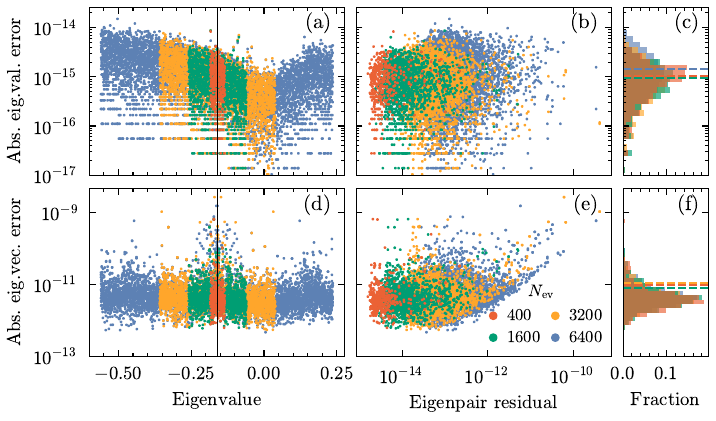}
    \caption{\textbf{Errors of the eigenvalues and eigenvectors obtained with POLFED for different numbers of requested eigenpairs.} The first row (panels (a)--(c)) and the second row (panels (d)--(f)) show the absolute errors of the eigenvalues, $|\varepsilon_i - E_i|$, and of the eigenvectors, $\|\psi_i - u_i\|$, respectively. Colors distinguish the number of requested eigenpairs $N_{\rm ev}$, as indicated in the legend. In the first column (panels (a) and (d)) the errors are plotted as a function of the eigenvalue, with the solid vertical line marking the target energy of the run. The second column (panels (b) and (e)) shows them as a function of the eigenpair residual $\|H\psi_i - \varepsilon_i\psi_i\|$, and the third column (panels (c) and (f)) their distributions, with dashed lines indicating the mean values. Results are shown for the disordered XXZ model in the zero-magnetization sector with $L=18$ and block size $4$.}
\label{fig:error:howmany}
\end{figure}

As its name indicates, POLFED is an exact diagonalization method, so that its results are converged to a controlled accuracy. In all benchmarks and demonstrations presented above we required every returned eigenpair to satisfy the convergence criterion $\|H\psi_i - \varepsilon_i \psi_i\| \leq 10^{-9}$; we refer to this quantity as the eigenpair residual. In this section we quantify the remaining error with respect to full exact diagonalization. Let us note that the distinction between exact eigenpairs $E_i, u_i$ and their Ritz approximations $\varepsilon_i,\psi_i$ was disregarded in numerical sections of this work, but we bring this distinction back for this Appendix, as it focuses on comparison with full exact diagonalization.

In this work we use $\|\bullet\|$ for the standard $\mathrm{L}^2$ norm, $\|\psi\| = \sqrt{\sum_{j=1}^{\mathcal{D}} |\psi_j|^2}$,
where $\psi_j$ denotes the $j$-th component of $\psi$. We will consider a few different types of error. Firstly, for the eigenvalues we will use the conventional absolute and relative errors, defined as $|\varepsilon_i - E_i|$ and $|\varepsilon_i/E_i - 1|$, respectively. Proceeding to the eigenvectors, we again introduce the absolute and relative errors of the two vectors $\psi_i$ and $u_i$, defined as
\begin{align}
    \|\psi_i - u_i\| = \sqrt{\sum_{j=1}^{\mathcal{D}} \left|\psi_{j,i} - u_{j,i}\right|^2}
    \qquad
    \text{and}
    \qquad
    \|\psi_i - u_i\|_{\rm rel} = \sqrt{\sum_{j=1}^{\mathcal{D}} \left|\frac{\psi_{j,i}}{u_{j,i}} - 1\right|^2},
\end{align}
where $\psi_{j,i}$ and $u_{j,i}$ are the $j$-th components of $\psi_i$ and $u_i$. We also define the errors with respect to the single maximal coefficient, again in the absolute and relative form, $\max_j|\psi_{j,i} - u_{j,i}|$ and $\max_j|\psi_{j,i}/u_{j,i} - 1|$.

Figure~\ref{fig:error:eigenpair} shows how several indicators of the error behave for a typical POLFED run. All of them are well behaved. The eigenvalues are obtained essentially to machine precision, with absolute errors of the order of $10^{-15}$. The eigenvectors are likewise well converged: their absolute errors fluctuate around $10^{-11}$ and the relative errors around $10^{-6}$.

The width of the filter is controlled by the polynomial order and is adjusted to yield the requested number of eigenpairs. Increasing $N_{\rm ev}$ therefore lowers the polynomial order, and the transformed eigenvalues become more densely spaced, cf.\ Fig.~\ref{fig:EigenvalueTransformation}. Neighboring eigenvalues are then harder to resolve and the probability of artificial degeneracies increases, i.e., two non-degenerate eigenvalues being mapped onto the same transformed value. This is possible because the transformation is not injective: since the filter is peaked at the target energy $\tilde{\lambda}$, eigenvalues lying symmetrically on either side of it are mapped onto nearly the same value.
 
Figure~\ref{fig:error:howmany} shows how the number of requested eigenpairs affects the accuracy of the results. The eigenvalues remain accurate up to numerical precision, whereas the accuracy of the eigenvectors slowly degrades as $N_{\rm ev}$ increases. Figure~\ref{fig:error:howmany}(f) shows that the shape of the distribution is essentially independent of $N_{\rm ev}$; only its tails grow longer as more eigenpairs are sampled, and these tails shift the mean error of the run (dashed lines in the histograms).
 
In both Figs.~\ref{fig:error:eigenpair} and \ref{fig:error:howmany}, the eigenpair residuals lie well below the threshold value of $10^{-9}$ for all runs, with most of the computed eigenpairs yielding residuals several orders of magnitude smaller. For larger $N_{\rm ev}$, a negligible fraction of states approaches a residual of $10^{-10}$, as depicted in Fig.~\ref{fig:error:howmany}, but all remain below the threshold of $10^{-9}$.

\begin{figure}[b!]
    \centering
    \includegraphics[width=0.999\columnwidth]{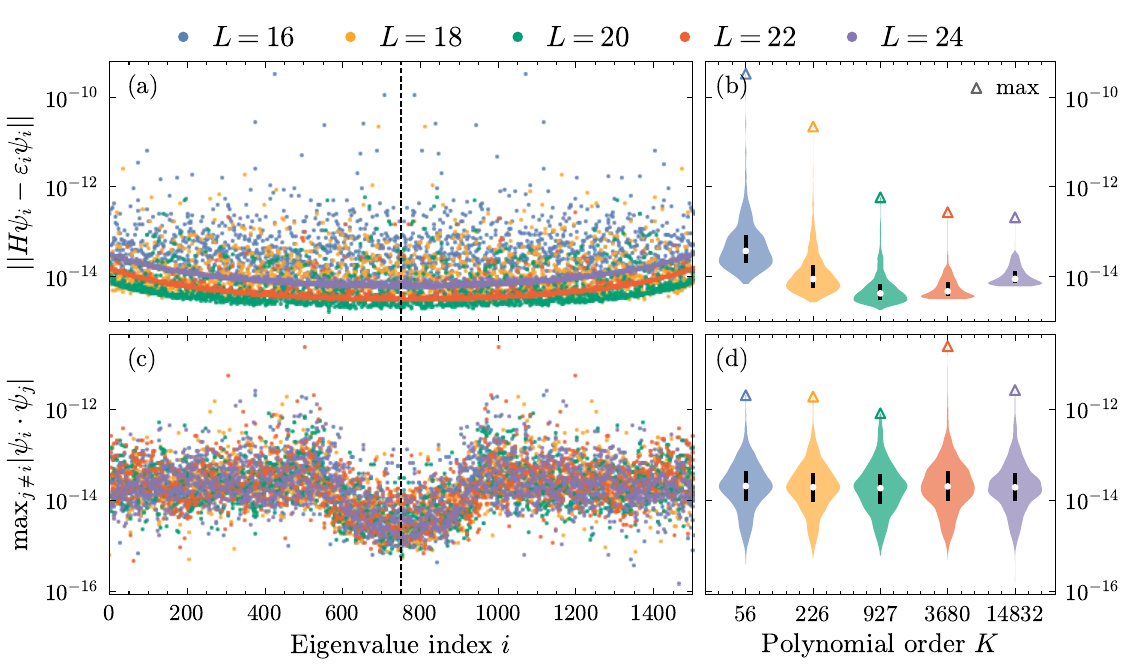}
    \caption{\textbf{Accuracy of POLFED eigenpairs for increasing system size and polynomial order of the filter.} The first row (panels (a) and (b)) shows the eigenpair residual $\|H\psi_i - \varepsilon_i\psi_i\|$ and the second row (panels (c) and (d)) shows the loss of orthogonality $\max_{j\neq i}|\psi_i\cdot\psi_j|$. In the first column (panels (a) and (c)) both quantities are plotted as a function of the eigenvalue index $i$, with the dashed vertical line marking the target energy. Since the $N_{\rm ev}$ eigenpairs closest to the target energy are returned, it lies in the middle of the index range for all $L$. The second column (panels (b) and (d)) shows their distributions as a function of $K$: violins represent the distributions, black bars the interquartile range, white dots the medians, and open triangles the maxima. Results are shown for the disordered XXZ model in the zero-magnetization sector, obtained with $N_{\rm ev}=1500$ requested eigenpairs and block size $16$.}
\label{fig:large:L:errors}
\end{figure}

Two of the $N_{\rm ev}$ requested eigenpairs may occasionally form an artificial degeneracy due to lack of injectivity of the spectral transformation $P_{\tilde{\lambda}}^K$: two eigenvalues $E_1,E_2$ may be transformed to similar values $P(E_1)\approx P(E_2)$, in which case they fail to fully converge and their residual exceeds the value, $\|H \psi_i - \varepsilon_i \psi_i\| > 10^{-9}$. In such a scenario, only these two eigenpairs are not fully converged, while the remaining $N_{\rm ev}-2$ requested eigenpairs are converged. The algorithm then returns only converged eigenpairs, with eigenvalues $E$ lying between the two artificial degeneracies $E\in [E_1,E_2]$, to ensure there are no eigenvalues missing in the spectrum. Although the probability of such an event is small, it does occur. For this reason we recommend that the user does not set this bound too strictly. Artificial degeneracies can also be recognized from the approximate symmetry of the eigenvector errors with respect to the solid vertical line marking the target energy in the first column of both figures. This symmetry is particularly clear in Fig.~\ref{fig:error:howmany}(d), where the larger errors appear in pairs.

Since full exact diagonalization is not feasible for larger system sizes, we use the eigenpair residual and the loss of orthogonality to show in Fig.~\ref{fig:large:L:errors} that the accuracy of the POLFED method persists to larger $L$. Moreover, polynomial orders $K$ that grow exponentially with system size do not damage the accuracy of the algorithm: the $K$ matrix multiplications within the Clenshaw algorithm remain numerically stable.

\section{The Kernel Polynomial Method}
\label{app:kpm}
The idea of the KPM is to expand an arbitrary function $f$ (can be scalar or matrix function) in terms of Chebyshev polynomials, and then evaluate it for some finite truncation of the expansion. We refer the reader to \cite{kpm} for a more detailed discussion about the method. 
    
Let us define two scalar products of functions $f$ and $g$ with the weight function $w(x)$
\begin{align}
    \label{eq:scalarproducts_a}
    \braket{f}{g}_{1} &= \int_a^b f(x) g(x) w(x) \text{d}x, \\
    \braket{f}{g}_{2} &= \int_a^b \frac{f(x) g(x)}{ w(x)} \text{d}x,\label{eq:scalarproducts_b}
\end{align}
where $w(x)$ is the Chebyshev weight function $w(x)=(1-x^2)^{-1/2}/\pi$, and the interval of integration is $[-1,1]$. For Chebyshev polynomials of the first kind defined by Eq.~\eqref{eq:chebyshev_definition}, orthogonality relations read:
\begin{align}
    \braket{T_n}{T_m}_1 = \frac{1+\delta_{n,0}}{2}\delta_{n,m}.
\end{align}
We can expand the function $f$ as:
\begin{align}
    f(x) = \sum_{n=0}^{\infty}\frac{\braket{f}{T_n}_1}{\braket{T_n}{T_n}_1} T_n(x)= \alpha_0 + 2 \sum_{n=1}^{\infty}\alpha_n T_n(x),
\end{align}
with the coefficients:
\begin{align}
    \alpha_n = \braket{f}{T_n}_1 = \int_{-1}^1 \frac{1}{\pi\sqrt{1-x^2}} f(x) T_n(x) \text{d}x,
\end{align}
so the calculation of the expansion coefficients relies on an integral weighted by $w(x)$. In practical applications involving matrix operators, explicitly handling this weight function introduces significant and avoidable computational overhead. This difficulty is circumvented by absorbing the weight function into a modified set of Chebyshev polynomials. Let us introduce the rescaled polynomials,
\begin{align}
    \phi_n(x) = \frac{T_n(x)}{\pi \sqrt{1-x^2}}.
\end{align}
These polynomials still obey the same orthogonality relation as Chebyshev polynomials, but with a second type of scalar product \eqref{eq:scalarproducts_b},
\begin{align}
    \braket{\phi_n}{\phi_m}_2 = \frac{1+\delta_{n,0}}{2}\delta_{n,m}.
\end{align}
Expansion of the function $f$ in Chebyshev polynomials now gives the expression:
\begin{align}
    f(x) = \frac{1}{\pi \sqrt{1-x^2}}\bigg[ \mu_0 + 2 \sum_{n=1}^{\infty} \mu_n T_n(x) \bigg],
\end{align}
with coefficients $\mu_n$, also called the (modified) moments, which are expressed in integral form,
\begin{align}
    \label{eq:final_moments}
    \mu_n = \int_{-1}^{1} f(x) T_n(x) \text{d}x.
\end{align}
    
When expanding a function that has singularities or discontinuous points, in Chebyshev polynomials (or other orthogonal polynomials), a very well-known phenomenon of Gibbs oscillations occurs. These oscillations arise when truncating an infinite series to a finite order, resulting in poor precision and fluctuations near points where the function is not continuously differentiable. Oscillations can be reduced (or even removed) with the use of kernels $f(x) \to f_{\rm KPM}(x)$, we modify the coefficients $\mu_n$ with some additional prefactor $g_n$ that decreases for higher-order contributions. Typical kernels used are Jackson, Lorentz, and Lanczos, to name a few; for our calculations of the density of states, Jackson's and Lanczos's are the most popular ones:
\begin{align}
    \label{eq:kernels_jackson_and_lorenz}
    g_n^J &= \frac{1}{N+1}\bigg[(N-n+1)\cos(\frac{\pi n}{N+1}) + \sin(\frac{\pi n}{N+1})\cot(\frac{\pi}{N+1})  \bigg] \\
    g_n^L &= \bigg( \frac{\sin(\pi n / N)}{\pi n / N} \bigg)^M,  \quad\quad\quad\quad\quad\quad\quad\quad\quad\quad\quad\quad\quad\quad \text{usually } M=3.
\end{align}
The modified expression of our function then reads: 
\begin{align}
    f_{\rm KPM}(x) = \frac{1}{\pi \sqrt{1-x^2}}\bigg[ \mu_0 g_0 + 2 \sum_{n=1}^{N-1} \mu_n g_n T_n(x) \bigg],
\end{align}
for moments $\mu_n$ \eqref{eq:final_moments} and the kernel $g_n$ \eqref{eq:kernels_jackson_and_lorenz}. For most applications, the Jackson kernel turns out to be the best option. 
Values of moments are now function-specific and can be of a different form. In practice, depending on the problem of our interest (function $f$), the calculation of moments usually reduces to evaluating an observable $A$ over arbitrary states $\ket{\alpha}, \ket{\beta}$, or taking a trace,
\begin{align}
    \mu_n = \bra{\alpha} A T_n(\tilde{H}) \ket{\beta}, \quad \quad \mu_n = \Tr \big\{ A T_n(\tilde{H}) \big\}.
\end{align}
Note that traces may be evaluated stochastically as:
\begin{align}
     \Tr \big\{ A T_n(\tilde{H}) \big\} = \frac{1}{R} \sum_{r=1}^{R} \bra{r} A T_n(\tilde{H}) \ket{r},
\end{align}
where $\ket{r}$ are random vectors, with vector coefficients distributed independently according to a standard normal distribution, without imposing normalization.

\section{Details of the polynomial order}
\label{app:details:on:the:polynomial:order}

In Sec.~\ref{sec:polynomial:spectral:filtering}, we introduced the polynomial spectral transformation with on-the-fly mapping, for which it is necessary to perform $K$ matrix multiplications at every step, arguing that this is the most time-demanding part of the problem. Later, in Sec.~\ref{sec:order:of:spectral:filtering}, we showed how we find this polynomial order, and in Sec.~\ref{sec:bench:parts} we demonstrated that, with the exception of large numbers of requested eigenpairs, the spectral transformation is truly the bottleneck of the algorithm.

In this Appendix, we demonstrate the representative values of the polynomial order $K$. We compare the polynomial order found exactly with bisection, as suggested in Sec.~\ref{sec:order:of:spectral:filtering}, to the estimate $K_{\rm estimate}$ from Eq.~\eqref{eq:K:approx}. In Table~\ref{tab:polynomial:order:values}, we list specific values for different system sizes; the agreement tends to be adequate. The table also supports the claim that the order scales linearly with the Hilbert-space dimension, as the ratio $K/\mathcal{D}$ is almost constant.

\begin{table}[htb!]
\centering
\begin{tabular}{l c c c c c c}
\hline\hline
\multicolumn{7}{c}{XXZ model} \\
\hline\hline
$L$ & 16 & 18 & 20 & 22 & 24 & 26 \\
\hline
$K$ & 57 & 231 & 937 & 3736 & 15024 & 60199 \\
$K_{\mathrm{estimate}}$ & 58 & 237 & 961 & 3830 & 15399 & 61704 \\
$K/\mathcal{D}$ & $4.4\times10^{-3}$ & $4.8\times10^{-3}$ & $5.1\times10^{-3}$ & $5.3\times10^{-3}$ & $5.6\times10^{-3}$ & $5.8\times10^{-3}$ \\
$\tilde{\lambda}$ & 0.076 & 0.062 & 0.027 & 0.055 & 0.069 & 0.059 \\
$\delta$ & $4.5\times10^{-2}$ & $1.1\times10^{-2}$ & $2.8\times10^{-3}$ & $6.9\times10^{-4}$ & $1.7\times10^{-4}$ & $4.3\times10^{-5}$ \\
\hline\hline
\end{tabular}
\caption{\textbf{Values of the polynomial order parameter for different system sizes.} Polynomial order $K$ for the XXZ model with $J=1$, $\Delta=0.55$, $W=1$ and different system sizes $L$ in the zero-magnetization sector, obtained with $N_{\rm ev}=1500$ requested eigenpairs at the peak of the density of states and the cutoff value $\Omega=0.17$. The value $K_{\rm estimate}$ is obtained from Eq.~\eqref{eq:K:approx}, $\tilde{\lambda}$ is the rescaled energy with the largest density of states, and $\delta$ represents the half-width of the targeted interval $[\tilde{\lambda}-\delta, \tilde{\lambda}+\delta]$, obtained numerically by bisection of Eq.~\eqref{eq:Nev:exact}.}
\label{tab:polynomial:order:values}
\end{table}

\begin{figure}[htb]
    \centering
    \includegraphics[width=0.99\columnwidth]{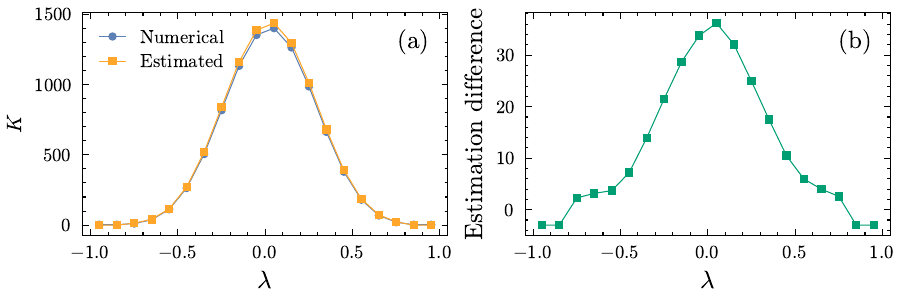}
    \caption{\textbf{Values of the polynomial order parameter for different targeted energies.} The figure shows the polynomial order $K$ for the XXZ model with $J=1$, $\Delta=0.55$, $W=1$ and $L=20$ in the zero-magnetization sector, obtained with $N_{\rm ev}=1000$ requested eigenpairs and the cutoff value $\Omega=0.17$. (a) Values of the polynomial order $K$ obtained numerically, as well as from the empirical estimate in Eq.~\eqref{eq:K:approx}, as a function of the rescaled targeted energy. (b) The difference between the estimated and the numerically obtained polynomial order.}
\label{fig:polynomial:order:target}
\end{figure}

Next, in Fig.~\ref{fig:polynomial:order:target} we also demonstrate good agreement of Eq.~\eqref{eq:K:approx} with the numerically obtained values of the order $K$ across the whole range of $\tilde{\lambda}$. The numerical value of $K$ is obtained by requiring that the polynomial filter is above truncation threshold, $P_{\tilde\lambda}^K(\tilde E)\geq \Omega$, precisely in the range containing $N_{\rm ev} $ eigenvalues, namely $\tilde E\in[\tilde \lambda-\delta,\tilde\lambda+\delta]$.

\section{Models}
\label{app:models}
For analyzing the POLFED method, we used several quantum many-body models with different connectivity structures. For spin-$1/2$ systems, we considered the disordered XXZ chain, the $J_1$--$J_2$ model, and its natural extension, the $J_1$--$J_2$--$J_3$ model. For fermions, we used a constrained version of the Sachdev-Ye-Kitaev (SYK) model. In addition, \inh{Polfed.jl} also implements the quantum random energy model (QREM) and the quantum sun model, which we define below for completeness.

We begin with the XXZ model:
\begin{align}
    \label{eq:xxz}
    H = J \sum_i^{L}\left(S^x_{i} S^x_{i+1}+ S^y_{i} S^y_{i+1} + \Delta S^z_{i}S^z_{i+1}\right) + \sum_i^L h_i S^z_i.
\end{align}
Here, $h_i$ are independent random fields uniformly distributed over the interval $[-W, W]$. In our calculations, we fixed the parameters to $W=1$, $J=1.22$, and $\Delta = 0.55$.

Next, we considered the $J_1$--$J_2$ model, which introduces additional off-diagonal matrix elements through next-nearest-neighbor couplings:
\begin{align}
    \label{eq:j1j2}
    H =
    J_1 \sum_{i=1}^L
    \left(
        S_i^x S_{i+1}^x
        + S_i^y S_{i+1}^y
        + \Delta_1 S_i^z S_{i+1}^z
    \right)
    +
    J_2 \sum_{i=1}^L
    \left(
        S_i^x S_{i+2}^x
        + S_i^y S_{i+2}^y
        + \Delta_2 S_i^z S_{i+2}^z
    \right)
    +
    \sum_{i=1}^L h_i S_i^z .
\end{align}
As before, $h_i$ are uniformly distributed over the interval $[-W,W]$. The parameters were set to $W=1$, $J_1=1$, $J_2=0.5$, $\Delta_1=1$ and $\Delta_2=1$.

To further test the scaling with the number of off-diagonal matrix elements, we also included the $J_1$--$J_2$--$J_3$ model:
\begin{align}
    \label{eq:j1j2j3}
    H =
    &J_1 \sum_{i=1}^L
    \left(
        S_i^x S_{i+1}^x
        + S_i^y S_{i+1}^y
        + \Delta_1 S_i^z S_{i+1}^z
    \right)
    +
    J_2 \sum_{i=1}^L
    \left(
        S_i^x S_{i+2}^x
        + S_i^y S_{i+2}^y
        + \Delta_2 S_i^z S_{i+2}^z
    \right)\\ \notag
    +
    &J_3 \sum_{i=1}^L
    \left(
        S_i^x S_{i+3}^x
        + S_i^y S_{i+3}^y
        + \Delta_3 S_i^z S_{i+3}^z
    \right)
    +
    \sum_{i=1}^L h_i S_i^z .
\end{align}
We used the same parameters as in the $J_1$--$J_2$ model, with the additional coupling $J_3=0.225$ and $\Delta_3=1$.

To extend the analysis to fermionic systems, we considered the SYK model~\cite{Sachdev93syk, Maldacena16syk} with two-particle interactions:
\begin{align}
    \label{eq:syk4_d}
    H = \sum_{\hat{d}(i,j,k,l)\leq d} U_{ijkl} c^\dagger_i c^\dagger_j c_k c_l.
\end{align}
The natural formulation of the four-point SYK model involves all-to-all interactions. Here, however, we use a constrained version in which the parameter $d$ specifies the maximal support of an interaction term. The constraint is defined by
\begin{align}
    \hat{d}(i,j,k,l) = \max\{i,j,k,l\} - \min\{i,j,k,l\} + 1 \leq d.
\end{align}
We denote the SYK model with two-particle interactions subject to this maximal support restriction as $\text{SYK}_4^d$. The couplings $U_{ijkl}$ satisfy the symmetry relations required by hermiticity and fermionic antisymmetry:
\begin{align}
    U_{ijkl} = -U_{jikl} = -U_{ijlk} = U_{klij}^{\ast}, \qquad \overline{U_{ijkl}} = 0, \qquad \overline{|U_{ijkl}|^2} = U^2.
\end{align}
Thus, interactions are allowed only within a finite support $d$. Defining the number operator $n_i=c_i^\dagger c_i$, the model can be rewritten in the form:
\begin{align}
    \notag
    H &= 2 \sum_{\substack{x(i,j)\\\hat{d}(i,j)\leq d}} U_{ij,ji} n_i n_j + 4 \sum_{\substack{x(i,j,k)\\\hat{d}(i,j,k)\leq d}} U_{ij,ki} n_i c^{\dagger}_j c_k + \sum_{\substack{x(i,j,k,l)\\ \hat{d}(i,j,k,l)\leq d}} U_{ij,kl}c_i^{\dagger} c_j^{\dagger} c_k c_l,
\end{align}
where $x(i,j,k,l)$ denotes mutually distinct indices. For illustration, for $d=3$ and arbitrary system size $L$ one obtains
\begin{align}
    H = \sum_{i=1}^{L} 2&(U_{i,i-2,i-2,i}n_{i} n_{i-2} + U_{i,i-1,i-1,i}n_{i} n_{i-1}) \\
    + 4&(U_{i,i-2,i-1,i}n_{i}c_{i-2}^{\dagger}c_{i-1} + U_{i,i-1,i+1,i}n_{i}c_{i-1}^{\dagger}c_{i+1} + U_{i,i+1,i+2,i}n_{i}c_{i+1}^{\dagger}c_{i+2}) + h.c.
\end{align}

Another model we tested is the quantum random energy model (QREM), which augments the classical random energy model~\cite{derrida_rem_model} by a transverse field~\cite{qrem_mobility_edge_14,qrem_mbl_baldwin_2016}. Its Hamiltonian reads:
\begin{align}
    \label{eq:qrem}
    H = \sum_{\alpha} E_{\alpha}\ket{\alpha}\bra{\alpha} - \Gamma \sum_{i=1}^{L} \sigma_i^x.
\end{align}
Here, $\{\ket{\alpha}\}$ denotes the computational basis of spin configurations, $E_{\alpha}$ are independent Gaussian random energies assigned to each basis state (with mean zero and standard deviation $\sigma = \sqrt{L/2}$), and $\Gamma$ is the transverse-field strength, with $\sigma_i^x$ denoting the standard Pauli matrix. The first term is diagonal in the computational basis and encodes an uncorrelated random energy landscape, while the transverse field introduces off-diagonal transitions between basis states differing by a single spin flip. The QREM provides a simple mean-field model for studying many-body localization and ergodicity breaking~\cite{qrem_mobility_edge_14,qrem_mbl_baldwin_2016,fsa_mbl_pietracaprina}.

\begin{figure}[htb!]
    \centering
    \includegraphics[width=0.65\textwidth]{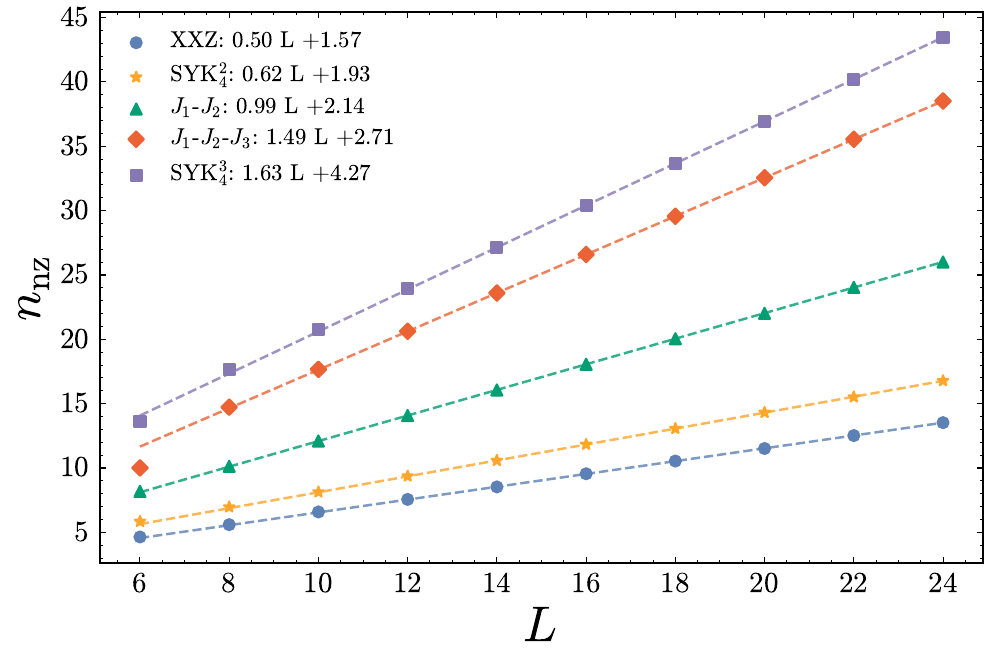}
    \caption{\textbf{Number of nonzero off-diagonal matrix elements per row.} Number of nonzero off-diagonal matrix elements per row as a function of system size $L$ for the XXZ model in Eq.~\eqref{eq:xxz}, the $J_1$--$J_2$ model in Eq.~\eqref{eq:j1j2}, the $J_1$--$J_2$--$J_3$ model in Eq.~\eqref{eq:j1j2j3}, and the constrained $\text{SYK}_4^d$ model in Eq.~\eqref{eq:syk4_d}.}
    \label{fig:scalling:of:offdiagonals}
\end{figure}

For our analysis, it is important to track how the number of off-diagonal matrix elements scales with increasing system size. Since the Hilbert-space dimension $\mathcal{D}$ grows exponentially, the total number of off-diagonal matrix elements $N_{\rm off}$ also grows exponentially. We therefore focus on the average number of off-diagonal matrix elements per row, $n_\mathrm{nz} = N_{\rm off}/\mathcal{D}$. This scaling is shown in Fig.~\ref{fig:scalling:of:offdiagonals}.

To quantify the sparsity of these matrices, $n_\mathrm{nz}$ can be derived analytically using combinatorics. For periodic boundary conditions, with the spin models in the zero-magnetization sector and the $\mathrm{SYK}_4^d$ model at half filling ($N_p = L/2$), the exact closed-form expressions and their large-$L$ asymptotic limits are:
\begin{align}
    n_\mathrm{nz}^{\mathrm{XXZ}} &= \frac{L^2}{2(L-1)} \approx \frac{1}{2}L\,, \\
    n_\mathrm{nz}^{J_1\text{--}J_2} &= \frac{L^2}{L-1} \approx L\,, \\
    n_\mathrm{nz}^{J_1\text{--}J_2\text{--}J_3} &= \frac{3L^2}{2(L-1)} \approx \frac{3}{2}L\,,\\
    n_\mathrm{nz}^{\mathrm{SYK}_4^d} &\approx \left[\frac{3}{8}\binom{d}{3} + \frac{d}{2} - \frac{1}{2^{d-1}} + \frac{1}{2^{2d-1}}\right]L\,.
\end{align}
The exact closed-form expression for the $\mathrm{SYK}_4^d$ model is omitted for brevity. We note the excellent agreement between these asymptotic formulas and the exact data for numerically relevant system sizes, presented in Fig.~\ref{fig:scalling:of:offdiagonals}.

\subsection{Quantum Sun models}
\label{app:qsun}
Finally, \inh{Polfed.jl} also implements the quantum sun model~\cite{Suntajs22zeroDim}, a toy model of many-body ergodicity breaking. It describes $L$ localized spins coupled to a small ergodic quantum dot via spatially decaying interaction (``rays'')~\cite{Suntajs22zeroDim}. The Hamiltonian reads
\begin{align}
    \label{eq:qsun}
    H = H_{\mathrm{dot}} + g_0\sum_{j=1}^{L} \alpha^{u_j}\, S^x_{n(j)} S^x_j + \sum_{j=1}^{L} h_j\, S^z_j,
\end{align}
where $H_{\mathrm{dot}}$ is proportional to $2^N \times 2^N$ random matrix drawn from the Gaussian orthogonal ensemble (GOE), representing all-to-all interactions within an ergodic quantum dot of $N$ spins. Each outer spin $j$ is coupled to a randomly selected dot spin $n(j)$ with strength $g_0\alpha^{u_j}$, where $\alpha$ is the control parameter and $u_j \propto j$ are random exponents drawn from a uniform distribution around $j-1$ with width $2\zeta$. The fields $h_j$ are independent random variables from a uniform distribution with mean $h$ and width $2W$. The parameter $\alpha$ tunes the system across an ergodicity-breaking transition; based on theoretical arguments, for $\alpha > \alpha_c = 1/\sqrt{2}$ the dot thermalizes the entire system, whereas for $\alpha < \alpha_c$ the avalanche stalls and ergodicity is broken~\cite{Suntajs24ultrametric}.

We follow the normalization of $H_{\mathrm{dot}}$ introduced in Ref.~\cite{Suntajs24ultrametric} and subsequently used in Refs.~\cite{Kliczkowski24fading,Swietek25fading,Swietek25scaling}, i.e., we set the Hilbert-Schmidt norm of random matrix to unity, $||R_{\mathrm{dot}}||=1$, where $H_{\mathrm{dot}}=\gamma R_{\mathrm{dot}}$ and $||O||^{2}={\rm Tr}(O^2)/\mathcal{D}$. In practice, we first draw a $2^N \times 2^N$ matrix $A$ whose entries are independent normal random variables, and construct the symmetric matrix $R'_\mathrm{dot}=(A+A^\dagger)/\sqrt{2}$. Subsequently, we normalize $R'_\mathrm{dot}$ by factor $\sqrt{2^{N}+1}$ and embed it into $2^{N+L} \times 2^{N+L}$ matrix, i.e., ${R}_\mathrm{dot} = (R'_\mathrm{dot}\otimes \mathbb{I})/\sqrt{2^N+1}$, and ${R}_\mathrm{dot}$ is a $2^{N+L} \times 2^{N+L}$ block-diagonal matrix. The normalization factor ensures that $||R_{\mathrm{dot}}||=1$ for large $N$. Alternatively, $R_{\mathrm{dot}}$ could be normalized numerically.

We also implement the $U(1)$ symmetric variant of the quantum sun model, introduced in Ref.~\cite{Pawlik24mobilityEdge}, further modified with the canonical normalization condition of Ref.~\cite{Suntajs24ultrametric}, for which the phase transition point was established in Ref.~\cite{andreoni2026}. It is given by
\begin{align}
    \label{eq:qsun:U1}
    H = H_{\mathrm{dot}}^{U(1)} + g_0\sum_{j=1}^{L} \alpha^{u_j}\, \qty(S^x_{n(j)} S^x_j + S^y_{n(j)} S^y_j) + \sum_{j=1}^{L} h_j\, S^z_j,
\end{align}
where the interaction between the dot spins and the outer spins is $U(1)$ symmetric by the addition of $S^yS^y$ terms to the original interaction in Eq.~\eqref{eq:qsun}. Here, $H_{\mathrm{dot}}^{U(1)}$ represents all-to-all interactions within an ergodic quantum dot of $N$ spins that fulfill the  $U(1)$ symmetry. We adopt to $H_{\mathrm{dot}}^{U(1)}$ the canonical normalization condition as in Ref.~\cite{Suntajs24ultrametric}, i.e., $||H_{\mathrm{dot}}^{U(1)}||=\gamma$, where, however, the Hilbert-Schmidt norm $||O||^{2}={\rm Tr}(O^2)/\mathcal{D}^{U(1)}$ is related to Hilbert space of dimension $\mathcal{D}^{U(1)}$, corresponding to a chosen magnetization sector.

In practice, we first generate a matrix ${R}_{\mathrm{dot}}^{U(1),\mathrm{all}}$, which is done as follows. We start with the matrix $R'_\mathrm{dot}$, i.e., the same matrix as discussed above. Subsequently, we set to zero the elements of this matrix that correspond to the processes between the basis states of the dot that do not conserve the $U(1)$ symmetry, i.e., $R'_\mathrm{dot} \rightarrow {R'}^{U(1)}_\mathrm{dot}$. Finally, we then embed ${R'}^{U(1)}_\mathrm{dot}$ to the full $2^{N+L} \times 2^{N+L}$ matrix,
 i.e., ${R}_{\mathrm{dot}}^{U(1),\mathrm{all}} = {R'}^{U(1)}_\mathrm{dot}\otimes \mathbb{I}$. The dimension of matrix ${R}_{\mathrm{dot}}^{U(1),\mathrm{all}}$ is the sum of dimensions of Hilbert spaces of all magnetization sectors.

As the system is $U(1)$ symmetric, the matrix ${R}_{\mathrm{dot}}^{U(1),\mathrm{all}}$ is block-diagonal with respect to the magnetization sectors. Thus, after selecting a magnetization sector, we reduce the operator ${R}_{\mathrm{dot}}^{U(1),\mathrm{all}}$ to the corresponding Hilbert space of that sector,
${R}_{\mathrm{dot}}^{U(1),\mathrm{all}}   \rightarrow {R}_{\mathrm{dot}}^{U(1)}$. Here, $R_{\mathrm{dot}}^{U(1)}$ is an operator in the reduced Hilbert space of dimension $\mathcal{D}^{U(1)}$. Finally, we construct the dot operator with the normalized ${R}_{\mathrm{dot}}^{U(1)}$ as
\begin{equation}
H_{\mathrm{dot}}^{U(1)}  =\gamma \frac{{R}_{\mathrm{dot}}^{U(1)}}{\Big|\Big|{R}_{\mathrm{dot}}^{U(1)}\Big|\Big|},
\end{equation}
where the normalization is implemented numerically.

To give an example, for the zero-magnetization sector, the operator $H_{\mathrm{dot}}^{U(1)}$ is a matrix of size $\binom{L}{L/2} \times\binom{L}{L/2}$. As the dimension of this magnetization sector scales exponentially, i.e., $\mathcal{D}^{U(1)}=\binom{L}{L/2}\propto 2^{L}/\sqrt{L}$, the system is expected to exhibit behavior similar to that of the system without $U(1)$ symmetry, namely, for $\alpha > \alpha_c = 1/\sqrt{2}$ the dot thermalizes the entire system, whereas for $\alpha < \alpha_c$ the ergodicity is broken.

The quantum sun models in Eqs.~\eqref{eq:qsun} and \eqref{eq:qsun:U1} implicitly assume a spin-1/2 system, i.e., a system of qubits. However, we provide generalization which extends these models to arbitrary spin $S$ (qudits). In this framework, the local operators are promoted to the standard spin-$S$ generators ${S}^\kappa$ (with $\kappa = x,y,z$), acting on the full many-body Hilbert space as
\begin{equation}
    {S}^\kappa_j = \underbrace{\mathds{1}\otimes\cdots}_{j-1\ \mathrm{times}} \otimes {S}^\kappa \otimes \underbrace{\cdots\otimes\mathds{1}}_{L-j\ \mathrm{times}}\,.
\end{equation}
Each local degree of freedom now spans a $(2S+1)$-dimensional Hilbert space, with basis states labeled by the spin projection quantum numbers $m_z = -S, -S+1, \dots, S-1, S$. In contrast to the spin-$\tfrac{1}{2}$ case, where operators reduce to Pauli matrices, the higher-spin operators are represented by $(2S+1)\times(2S+1)$ matrices obeying the $\mathfrak{su}(2)$ algebra.

A convenient representation of these operators is given in the standard $S^z$ eigenbasis $\{\ket{a}\}$, where $a=1,\dots,2S+1$ corresponds to $m_z = S+1-a$. Their matrix elements take the form (we set $\hbar=1$):
\begin{align}
   ({S}^x)_{ab} &= \frac{1}{2}\qty(\delta_{a,b+1}+\delta_{a+1,b})\sqrt{(S+1)(a+b-1)-ab}\,,
   \\
   ({S}^y)_{ab} &= \frac{i}{2}\qty(\delta_{a,b+1}-\delta_{a+1,b})\sqrt{(S+1)(a+b-1)-ab}\,,
   \\
   ({S}^z)_{ab} &= (S+1-a)\,\delta_{a,b}\,,
\end{align}
or equivalently, these expressions follow from the ladder operator structure ${S}^\pm = {S}^x \pm i {S}^y$, with
\begin{equation}
    {S}^\pm \ket{m} = \sqrt{S(S+1) - m(m \pm 1)}\,\ket{m \pm 1}\,,
\end{equation}
which highlights that, for $S>1/2$, each local operator connects multiple neighboring spin sectors rather than acting as a simple two-level flip.

Accordingly, the central dot Hamiltonian $H_{\mathrm{dot}}$ is embedded into the grain subspace of dimension $(2S+1)^N \times (2S+1)^N$, and extended to the full system via tensor products with identity operators on the remaining sites.

An important distinction from the $S=\tfrac{1}{2}$ case is the rapid growth of the Hilbert-space dimension. For a system consisting of $N+L$ sites, the total dimension is $\mathcal{D} = (2S+1)^{N+L}$.
As a consequence, the value of ergodicity-breaking transition point might be influenced for higher spin $S$.
However, the full characterization of the transition with increasing $S$ is left for future work.

\section{Optimal number of requested eigenpairs $N_{\rm ev}$}
\label{app:optimal:number:of:nev}

    As discussed in Sec.~\ref{sec:bench:nev}, the runtime of POLFED depends on the requested number of eigenpairs $N_{\rm ev}$, as demonstrated in Fig.~\ref{fig:tcpu:Nev}. This turns out to be the only algorithm-related parameter that can be tuned to improve performance. In this Appendix, we elaborate on its optimal value.

    Which part of the program dominates the runtime depends on the regime of $N_{\rm ev}$. For small $N_{\rm ev}$, the spectral transformation is the most time-consuming part, as indicated in Fig.~\ref{fig:tcpu:parts}. A small number of eigenvalues lie in a very narrow interval $[\tl-\delta,\tl+\delta]$, whose width decreases exponentially with the system size (in the middle of the spectrum it is approximately proportional to $N_{\rm ev}/\mathcal{D}$). Resolving such an interval requires an exponentially large polynomial order $K$. In the opposite limit of large $N_{\rm ev}$, the number of Lanczos iterations $m$ required for convergence also becomes large. At the end of the program, the algorithm has to diagonalize the Lanczos matrix of dimension $ms\times ms$; when $m$ is sufficiently large, this final diagonalization becomes the bottleneck.

    As $N_{\rm ev}$ is increased, starting from the regime dominated by the spectral transformation, the order $K$ decreases (the spectral transformation becomes cheaper), while the diagonalization of the Lanczos matrix becomes more expensive. Consequently, there exists an optimal value of $N_{\rm ev}$ that minimizes the runtime, as evident from Fig.~\ref{fig:tcpu:Nev}.

    To estimate this optimum, we model the two dominant contributions. The total time spent on matrix multiplications during the whole algorithm is
    \begin{equation}
        \tmap = m\,K\,s\,\mathcal{D}\,\tzm ,
    \end{equation}
    where $m$ is the number of iterations needed for convergence, $s$ is the block size, $\mathcal{D}$ is the Hilbert-space dimension, and $\tzm$ is a prefactor that depends on the model (the structure of the Hamiltonian) and on the CPU architecture. The time needed for the final diagonalization is
    \begin{equation}
        \tdiag = (m\,s)^{l}\,\tzd ,
    \end{equation}
    where $l=2$ for the standard Lanczos factorization ($s=1$, tridiagonal matrix) and $l=3$ for the block Lanczos factorization ($s>1$). The dependence of $m$ and $K$ on $N_{\rm ev}$ is
    \begin{equation}
        m(N_{\rm ev}/s)=a+b\,N_{\rm ev}/s ,
        \qquad
        K=\frac{c\,\rho(\tl)\sqrt{1-\tl^{2}}}{N_{\rm ev}} .
        \label{eq:mK}
    \end{equation}
    The optimal value $N_{\rm ev}^{\ast}$ follows from minimizing the total time,
    \begin{equation}
        t = \tmap+\tdiag, \qquad \frac{\ddd t}{\ddd N_{\rm ev}}=0 .
        \label{eq:solving:for:optimal:nev}
    \end{equation}
    Other contributions to the runtime are neglected; the most important of them is the reorthogonalization, whose cost scales as $\mathcal{D}\,(ms)^{2}$.

    Inserting Eq.~\eqref{eq:mK} into Eq.~\eqref{eq:solving:for:optimal:nev} gives the stationarity condition,
    \begin{equation}
        N_{\rm ev}^{2}\,(N_{\rm ev}+\kappa)^{\,l-1} = R_{l} ,
    \end{equation}
    which is a cubic equation for $l=2$, and a quartic equation for $l=3$. The unique positive solutions for $l=2$ and $l=3$ are, respectively,
    \begin{equation}
        N_{\mathrm{ev}}^{*}\big|_{l=2} = \frac{\kappa}{3}\left[\,2\,\mathcal{C}\!\left(\frac{27R_{2}}{2\kappa^{3}}-1\right)-1\right] ,
        \qquad\qquad
        N_{\mathrm{ev}}^{*}\big|_{l=3} = \frac{\sqrt{\kappa^{2}+4\sqrt{R_{3}}}-\kappa}{2} ,
    \end{equation}
    with the constants
    \begin{equation}
        \kappa \equiv \frac{a\,s}{b},
        \qquad
        R_{l} \equiv \frac{A\,a\,s}{l\,\tzd\,b^{\,l}},
        \qquad
        A \equiv c\,\rho(\tl)\sqrt{1-\tl^{2}}\,\mathcal{D}\,\tzm ,
    \end{equation}
    and
    \begin{equation}
        \mathcal{C}(w)=
        \begin{cases}
            \cos\!\left[\tfrac{1}{3}\arccos w\right], & -1\le w\le 1,\\[4pt]
            \cosh\!\left[\tfrac{1}{3}\operatorname{arcosh} w\right], & w>1,
        \end{cases}
    \end{equation}
    i.e., $\mathcal{C}$ is the root $\mathcal{C}\ge\tfrac12$ of $4\mathcal{C}^{3}-3\mathcal{C}=w$.

    The obtained result should be regarded as practical guidance rather than a universal prediction. It depends on the block size $s$, the system size, the targeted part of the spectrum, and the model itself (its density of states and the structure of the Hamiltonian, which determines the cost of the matrix-vector multiplication). More importantly, $\tzm$ and $\tzd$ are not properties of the algorithm at all: they vary strongly with the hardware and the number of threads. In our experience, the dense diagonalization scales considerably worse with the thread count than the matrix-vector multiplications do. The estimate also neglects the $\mathcal{O}(\mathcal{D}\,(ms)^{2})$ cost of reorthogonalization, which may be relevant precisely in the crossover region where the two contributions compete, and it ignores the memory limit, which in practice often caps $N_{\rm ev}$ before the runtime optimum is reached.

\bibliography{literature.bib}

\end{document}

%% file: jlcodeblock_setup.tex
\usepackage[T1]{fontenc}
\usepackage{xcolor}
\usepackage{listings}
\usepackage{iftex}
\usepackage{jlcode}
\usepackage{hyperref}
\usepackage[most]{tcolorbox}

\makeatletter

\providecommand{\codeurl}{}
\providecommand{\codelinktext}{[Click here to download the code]}

\lstdefinestyle{code}{
  language=Julia,
  showstringspaces=false,
  keywordstyle={[1]\color[RGB]{255,134,2}},
  keywordstyle={[2]\color{jlliteral}},
  keywordstyle={[3]\color[RGB]{0,143,0}},
  keywordstyle={[4]\color{violet}},
  keywordstyle={[5]\color[RGB]{0,84,232}},
  commentstyle={\color{jlcomment}},
  stringstyle={\color[RGB]{216,72,0}},
  identifierstyle={\color{jlbase}},
  columns=fullflexible,
  keepspaces=true,
  upquote=true,
  frame=none
}

\lstdefinestyle{codenohighlight}{
  language=Julia,
  showstringspaces=false,
  keywordstyle={[1]\color{black}},
  keywordstyle={[2]\color{black}},
  keywordstyle={[3]\color{black}},
  keywordstyle={[4]\color{black}},
  keywordstyle={[5]\color{black}},
  commentstyle={\color{black}},
  stringstyle={\color{black}},
  identifierstyle={\color{black}},
  columns=fullflexible,
  keepspaces=true,
  upquote=true,
  frame=none
}

\newtcblisting{jlcodeblock}{
  enhanced,
  listing only,
  colback=gray!10,
  colframe=black!50,
  boxrule=0.25pt,
  arc=1.5mm,
  outer arc=1.5mm,
  boxsep=0pt,
  left=1.5mm,
  right=1.5mm,
  top=0.2mm,
  bottom=0.2mm,
  before skip=10pt,
  after skip=10pt,
  listing options={
    style=code,
    xleftmargin=0pt,
    xrightmargin=0pt,
    aboveskip=0pt,
    belowskip=0pt
  }
}

\newtcblisting{jlcodeblocknohighlight}{
  enhanced,
  listing only,
  colback=gray!10,
  colbacklisting=gray!10,
  colframe=gray!10,
  boxrule=0pt,
  arc=1.5mm,
  outer arc=1.5mm,
  sharp corners=none,
  boxsep=0pt,
  left=1mm,
  right=1mm,
  top=0.2mm,
  bottom=0.2mm,
  before skip=6pt,
  after skip=6pt,
  listing options={
    style=codenohighlight,
    xleftmargin=0pt,
    xrightmargin=0pt,
    aboveskip=0pt,
    belowskip=0pt
  }
}

\newcommand{\codelink}[1]{%
  \lstinputlisting[style=code]{#1}
  \noindent\begin{center}
  \filename@parse{#1}%
  \href{\codeurl/\filename@base.\filename@ext}{\textcolor{blue}{\codelinktext}}
  \end{center}
}

\newcommand{\linkonly}[1]{%
  \filename@parse{#1}%
  \href{\codeurl/\filename@base.\filename@ext}{\textcolor{blue}{\ttfamily[\filename@base.\filename@ext]}}
}

\newcommand{\jlinlcustomnohighlight}[1]{%
  \settowidth{\jlinlem}{\jlinlfont{m}}%
  \lstinline[language=Julia,style=codenohighlight,basicstyle=\jlinlfont,basewidth=\jlinlem]^^a7#1^^a7%
}
\newcommand{\inlinenohighlight}{\jlinlcustomnohighlight}

\makeatother